\documentclass[trackchanges]{aastex701}
\begin{document}

\title{Scattering, Migration, Re-circularization and Relaxation to Build Out Galaxy Disks with Exponential Profiles}
\author[orcid=0000-0002-6490-2156,gname=Curtis,sname=Struck]{Curtis Struck}
\affiliation{Iowa State University}
\email[show]{curt@iastate.edu}  
\author[orcid=0000-0002-1723-6330,gname=Bruce G.,sname=Elmegreen]{Bruce G. Elmegreen}
\affiliation{Retired, Katonah, NY 10536}
\email[show]{belmegreen@gmail.com}  
\author[orcid=0000-0003-2676-8344,gname= Elena,sname=D'Onghia]{Elena D'Onghia}
\affiliation{Department of Physics, University of Wisconsin-Madison, Madison, WI 53706, USA}
\affiliation{Department of Astronomy, University of Wisconsin-Madison, Madison, WI 53706, USA}
\email[show]{edonghia@astro.wisc.edu}  
\begin{abstract}

Scattering of stars by interstellar clouds or massive clumps increases the stellar velocity dispersion and promotes a radial disk profile that is exponential. Here we show that such scattering reaches a steady-state distribution function of stellar eccentricity, after which eccentricity increases and decreases occur at equal rates. The implication is that clump/cloud scattering recircularizes eccentric stellar orbits, keeping the stellar velocity dispersion in a limited range. This re-circularization regulates disk heating and maintains kinematic coherence, contributing to the longevity of disk structures.  The eccentricity distribution function and the presence of recircularizing cloud-star interactions are independent of cloud mass but the timescale to reach equilibrium decreases with increasing  mass. The calculations are made in the simplest possible disk system to highlight the effects of scattering without contamination from spiral waves, star formation, and other processes. The calculations also reveal a bifurcation in the disk evolutions whereby in a minority of cases temporary asymmetries in the clump spatial distribution drive the disks to an end state of increased velocity dispersion and orbital eccentricity corresponding to early type disks. Overall the models emphasize an important physical process that can make and maintain an exponential stellar disk in all galaxies with a cloudy interstellar medium.
\end{abstract}
\section{Introduction}
The radial profiles of galaxy stellar disks are approximately exponential for several scale lengths in the main parts, with either a continuation (Type I), a downturn (Type II), or an upturn (Type III) to a further exponential over many more scale lengths in the outer parts \citep{freeman70,erwin05,pohlen06,erwin08,meert15}.  Exponential profiles are observed in both spirals and dwarfs \citep{herrmann13}, whether barred or non-barred \citep{hunter06}, so theoretical processes involving bars \citep{foyle08,debattista06}, spiral waves \citep{roskar08a,berrier15}, or shear \citep{lin87,ferguson01,wang22} are not solely responsible for this structure. 
Deep JWST observations now reveal exponential-like stellar mass profiles in galaxies at z$\sim 2$ (Ferreira et al. 2022; Nelson et al. 2023), suggesting that the mechanisms responsible for exponential disks must operate early in cosmic history or be supplemented by more universal processes like stochastic scattering.


An example of a disk that would be difficult to explain by these models is a low-mass ($\sim10^8\;M_\odot$ in stars), low metallicity \citep[$<0.1$ solar;][]{morales11}, ``tadpole''  \citep{elmegreen12,putko19} that has no bar or spiral, and has a slowly rising rotation curve \citep{sanchez13} or single-peak HI line profile \citep{filho13} indicating weak shear. Like giant spirals, these tiny galaxies also have exponential or piece-wise exponentials in their underlying disks, sometimes out to 6 scale lengths \citep[e.g., the galaxy Kiso 3193, see][]{elmegreen12,putko25}. They also have one or more clumps of bright star formation, which is their only obvious structure.  
The clumpy morphologies and high velocity dispersions seen in early disk galaxies (Förster Schreiber et al. 2009; Genzel et al. 2011) reinforce the plausibility of gravitational scattering as a formative process, acting alongside or in place of spiral arm-driven migration in dynamically hot disks.

Numerical models with clumps like these quickly make exponentials \citep{bournaud07}, so it seems plausible that point-like scattering by interstellar clouds \citep{elmegreen13} or holes \citep{struck17} can make an exponential with a rate that depends on the cloud mass (or hole mass deficit). Support for this concept comes from purely stochastic models of scattering in a disk, which produce a profile $\exp(r/r_D)/r$ for radius $r$, with scale length $r_D=0.5\lambda/(1-2p)$,  mean free path $\lambda$, and probability $p$ for scattering in the outward direction \citep{elmegreen16}. This is an equilibrium profile if the net scattering is inward, $p<0.5$.  \cite{wu23} found non-equilibrium solutions of the same type with outward scattering, i.e., $p>0.5$.  Beyond the inner scale length, this profile is indistinguishable from an exponential. 
\cite{schonrich09} consider stochastic scattering at spiral corotation resonances to explain metallicity mixing, but do not derive the resulting radial profile. 
Unlike migration at corotation, which conserves orbital circularity and cannot build exponential profiles alone (Vera-Ciro, et al. 2014) cloud scattering directly redistributes angular momentum and energy, naturally building exponential profiles in both the mass and velocity structure \citep{wu20}.


Exponential profiles are also expected from simple galaxy formation models \citep{mestel63,fall80,dalcanton97,dutton09}, although the number of disk scale lengths that are possible this way \citep[$\sim5$;][]{freeman70,efstathiou00} is much smaller than the number observed in many cases. For example, there are 10 continuous scale lengths in the spiral galaxies NGC 4123 \citep{weiner01}, NGC 300 \citep{bland05,vlajic09,hillis16,jang20b}, and M101 \citep{mihos13,vandokkum14,jang20a}, and 8 continuous scale lengths in the spirals NGC 5383 \citep{barton97}, NGC 2903 \citep{abraham17,vijay25}, M95 and M96 \citep{watkins14}, and also in the dIrrs NGC 4163 and IZw115 \citep{hunter11}, HoII \citep{bernard12}, and NGC 2403 \citep{barker12,williams13}. 
Although angular momentum conservation during gas infall (e.g., Fall \& Efstathiou 1980, also see \citealt{berrier15}) can set up exponential initial conditions, such models struggle to explain the observed number of scale lengths or the maintenance of the profile over cosmic time in the presence of dynamical heating.

Exponentials on both sides of a profile break are a challenge to explain as well. NGC 2841 has a Type III exponential with an outer disk going to 31 mag arcsec$^{-2}$ in the g band, which is $\sim5R_{25}$ \citep{zhang18}; stars dominate gas all the way out.  NGC 7793 has a Type II profile that extends for $\sim4$ scale lengths on either side of a break \citep{vlajic11,radburn12}. The break position in NGC 7793 is independent of stellar age, suggesting a dynamical origin \cite[as for NGC 4244;][]{dejong07}, and the outer part gets shallower and redder with age, suggesting stellar scattering from the inner to the outer regions over time \citep{radburn12}.  This galaxy is also flocculent in the main optical disk, so spiral waves or global modes are not active now, and it has a slowly rising rotation curve up to 120 km s$^{-1}$, so shear is relatively weak \citep{radburn12}. Many other examples with breaks between exponentials are in \cite{martin12}, \cite{staudaher19}, and \cite{xu24}.  

A large survey of local disk galaxies by \cite{zheng15} shows a majority with Type II profiles in the bluer passbands that change to Type I in the red or in mass surface density when corrected for radial color gradients \citep[see also][]{bakos08}. The average mass profile for $\sim700$ galaxies in \cite{zheng15} is a single exponential extending for $\sim9$ scale lengths. 

Outer-disk structure can also include long-lived tidal features \citep{atkinson13} and stellar spirals driven by interactions \citep{purcell11,struck11,donghia16}. These features could scatter stars even further out, as could gaseous spirals driven beyond the outer Lindblad resonance of main-disk waves \citep{khop15}. Tidal forces can truncate or distort an exponential disk, as for M51 \citep{watkins15}. Tidal forces in the outer regions can scatter stars in the inner disk, sometimes recircularizing eccentric orbits \citep{quillen09}. The resulting radial profiles in these cases are not known, but it is conceivable that outer-disk processes combine with cloud scattering to make an exponential or piecewise exponential in a future relaxed state.

The radial distribution of stars that are scattered by stellar spiral arms is close to exponential \citep{berrier16}. \cite{sellwood02} showed that spiral arms add angular momentum to stars inside corotation and remove angular momentum from stars outside corotation, causing their gyrocenters to switch sides (``churning'') without changing their energies. A similar behavior occurs for stars on either side of a point mass, such as an interstellar cloud, which also does not change the energy of the stars if the cloud mass is much larger than the star mass \citep{wu22}. Spiral waves have a different effect on stars at the Lindblad resonances, which scatter stars to larger energies by resonantly pumping their epicyclic motions.  While both spirals and massive clouds churn and scatter field stars, only the cloud model can do this uniformly everywhere if there are clouds everywhere, making a continuous exponential profile \citep[and references therein]{wu20}. Spiral modes may also churn stars for many scale lengths if the corotation radii hop around over time \citep{sellwood02}, but long exponential profiles from this mechanism have not been demonstrated. 
In contrast to time-variable resonant scattering by spiral patterns, which is confined to regions around corotation or Lindblad resonances, cloud scattering acts more uniformly and continuously across the disk (\citealt{elmegreen16}, \citealt{wu20}), offering a more universal route to exponential structure.

A problem with purely stellar scattering models is that dust, molecules, and star formation rates are exponential too \citep{wong02,munoz09,hunt15,gonzalez16,casasola17,wang19,wang22,lin24}. The HI gas is more flat in the inner regions, but it has an exponential profile in the outer regions that is about the same for all disk galaxies when normalized to the radius and surface density of $1\;M_\odot$ pc$^{-2}$ \citep{bigiel12,wang16}. The outer HI scale length is twice the inner scale length for stars. This observation of gas exponentials supports models where gas viscosity \citep[e.g.,][]{lin87} or magnetic fields \citep{wang22} cause exponentials through disk accretion followed by star formation, but stochastic radial mixing in the gas can form exponentials too. For example, interstellar blowout and gas recycling through the halo makes an exponential gas profile in a gaseous analogy to cloud-star scattering \citep{struck18}. Disk-halo recycling at the star formation rate also maintains an exponential in the star-forming gas \citep{elmegreen14}.  


High-resolution observations with JWST reveal that exponential disks exist in young, gas-rich galaxies at redshifts of 2 \citep{ferreira22, nelson23} and higher \citep{ono25}. These galaxies often host clumpy ISM structures, indicating an environment ripe for star-cloud scattering. There is also evidence that larger clumps can have a longer lifespan \citep{sok25}. 
Spiral features have also been observed in galaxy disks at quite high redshifts, and could also scatter stars \citep[see e.g.,][]{kuhn24, salcedo2025}. These young spirals also often contain large clumps. Although there are thus multiple scattering sources, this work will focus on experiments isolating the effects of clump scattering. The interplay between feedback, gas recycling, and frequent stellar scattering may offer a coherent explanation for the rapid establishment and persistence of exponential profiles even at early times.

The purpose of this paper is to present new models of cloud-star scattering. Like our previous models, they always make an exponential disk profile, but now we consider the stellar eccentricity distribution in more detail. The model and its relation to earlier work is described in Sec. 2. The generic evolution of the model disks is illustrated with two examples in Sec. 3. General trends across the models and parameter dependences are discussed in Sec. 4, and the results summarized in Sec. 5. 


\section{Model Description}
\subsection{Numerical Disk Model}
The results presented below are based on test particle model scattering disks, which are modified versions of the code used in \cite{elmegreen13} and subsequent papers in this series. The disk is initialized with 9759 test particles placed on circular orbits in the central x-y plane in a rigid halo gravitational potential. The initial surface density is constant out to a radius of 8.0 units to highlight the subsequent development of the exponential. The test particles are then given random vertical (z) positions in the interval $(-0.3, 0.3)$ units (see below), and random vertical velocities in the interval $(-1.5, 1.5)$ units to produce a disk of finite thickness.

An ensemble of 65 massive clumps are placed on 40 evenly spaced rings in the disk at random initial azimuths to scatter the test particle stars. Specifically, the first ring has a radius of 1.2 units and subsequent rings have a spacing of 0.15 units (units are defined below). These clumps are maintained in constant circular orbits at their initial radii throughout the model runs, providing a clean scenario for particle scattering without the complications of clump-clump interactions or dynamical friction. The clumps are treated as softened point masses with a small softening length (i.e., 0.05 units) and their gravitational attraction on the test particles is computed at each time step. The initial distribution of the clumps extends well beyond the initial distribution of the test particles, so there are always scattering sites available as the stars move out (Fig.~\ref{fig:f1}). 

Following \cite{elmegreen13} and \cite{struck18}, a power-law form is assumed for the radial acceleration, given by,

\begin{equation}
\label{eqa}
g(r) = \frac{-GM_H}{H^2}
\left(\frac{r + 0.01}{H} \right) ^{-\gamma},
\end{equation}

\noindent where $r$ is the
central radius in three dimensions, and the 0.01 is a softening constant. Throughout this paper we use $\gamma = 1.2$ (positive), which corresponds to a slightly declining rotation curve, with a circular
velocity that goes as $R^\frac{1- \gamma}{2}$, where $R$ is the projected radius
within the disc. The gravitational mass within a radius $r$ is $M(r) = M_H
(r/H)^{2-\gamma}$ where $M_H$ and $H$ are the halo mass and radius. 

There is also an effective acceleration in the vertical direction as in (\citealt{struck17}), it is given by, 

\begin{equation}
\label{eqb}
g_z(r) = -0.3GM_H H^2
\left( \frac{z}{H} \right) \left( 
\frac{\sigma_{zo}}{\sigma_z} \right)^2,
\end{equation}

\noindent where $z$ is the distance from the midplane, $\sigma_z$ is the mean vertical velocity distribution averaged over the whole disk, and $\sigma_{zo}$ is the initial value of that quantity. The value of $\sigma_z$ is recalculated every 5.0 time units. It increases by less than a factor of two in the runs described below. In \citep{struck17}, the acceleration $g_z$ is linear in $z$, and $\sigma_z$ is assumed constant.  

\subsection{Units}
We use dimensionless units, specifically we take, $H = 1$, and $GM_H = 1.0$, then $V_H^2 = GM_H/H = 1$, and $T_H = H/V_H = 1$. As per previous papers in the series we can adopt representative scalings for a normal and a dwarf type disk. For the normal disk these are: $H = 2$  kpc, $V_H = 50$ km s$^{-1}$, $T = 40$ Myr, and $M_H = 5.0 \times 10^{9}\ M_{\odot}$. This gives rotation periods at $R = 2$ and 10 kpc of 250 and 1460 Myr, respectively. (The addition of a massive disk, bulge or bar component could increase rotation velocities and decrease these times significantly, without similarly affecting the velocity dispersions discussed below.) For the dwarf disk these are: $H = 0.5$ kpc, $V_H = 50$ km s$^{-1}$, $T = 9.8$ Myr, and $M_H = 2.9 \times 10^{8}\ M_{\odot}$, for a rotation period at $R = 0.5$ kpc of 61 Myr.

\subsection{Precessing Ellipse Fitting}
Important results of this paper derive from analyzing the evolution of the disk plane orbital eccentricities of the stellar ensemble due to scattering, so a key issue is how to compute those eccentricities. As we shall see below, this issue is complicated by the fact that the eccentricities can change on less than an orbital timescale. Thus, computing them over a full orbital period, or even over a half period from pericenter to apocenter is not entirely relevant, nor computationally efficient. 

Instead we have chosen to estimate eccentricities and semi-major axes using the precessing ellipse (p-ellipse) approximation of \cite{struck06}. This approximation is quite accurate for orbits in power-law potentials \citep{struck15}, especially near flat rotation curve potentials, and the ellipse quantities can be easily estimated from the positions and velocities of a single orbital point. 

As per \cite{struck06}, Appendix B, orbital positions and velocities yield specific energies and angular momenta for the orbits, which can be expressed in terms of eccentricities and semi-major axes. Unfortunately, in contrast to simple ellipses, these expressions cannot be explicitly inverted to obtain the latter. They can be well approximated by polynomials and inverted numerically. We have obtained the following form for orbital eccentricity in the above halo potential,

\begin{equation}
\label{eqc}
e = -74.454023\epsilon^4 - 2605.3345\epsilon^3 
- 34186.759\epsilon^2 - 199367.31\epsilon - 435974.34,
\end{equation}

\noindent where $\epsilon$ is the specific energy. The number of terms and the high precision of the coefficients is required for accurate representation of very eccentric orbits. This approximation does become inaccurate for large negative values of the energy, i.e., tightly bound particles. In this limit the first term of the approximation dominates and the computed eccentricities are negative. Since this affects only a small number of particles in the inner disk we have not attempted to correct the problem.

For the semi-major axes we have, 

\begin{equation}
\label{eqd}
a = 0.5*(h^2/GM(H))^{0.5556}*((1+e)^{-0.6} + (1-e)^{-0.6},
\end{equation}

\noindent where $h$ is the specific angular momentum of the orbit, defined as the local radius times the azimuthal velocity. 

\section{A fiducial model pair}
\subsection{Model Evolution}
\begin{figure*}[ht!]
\plotone{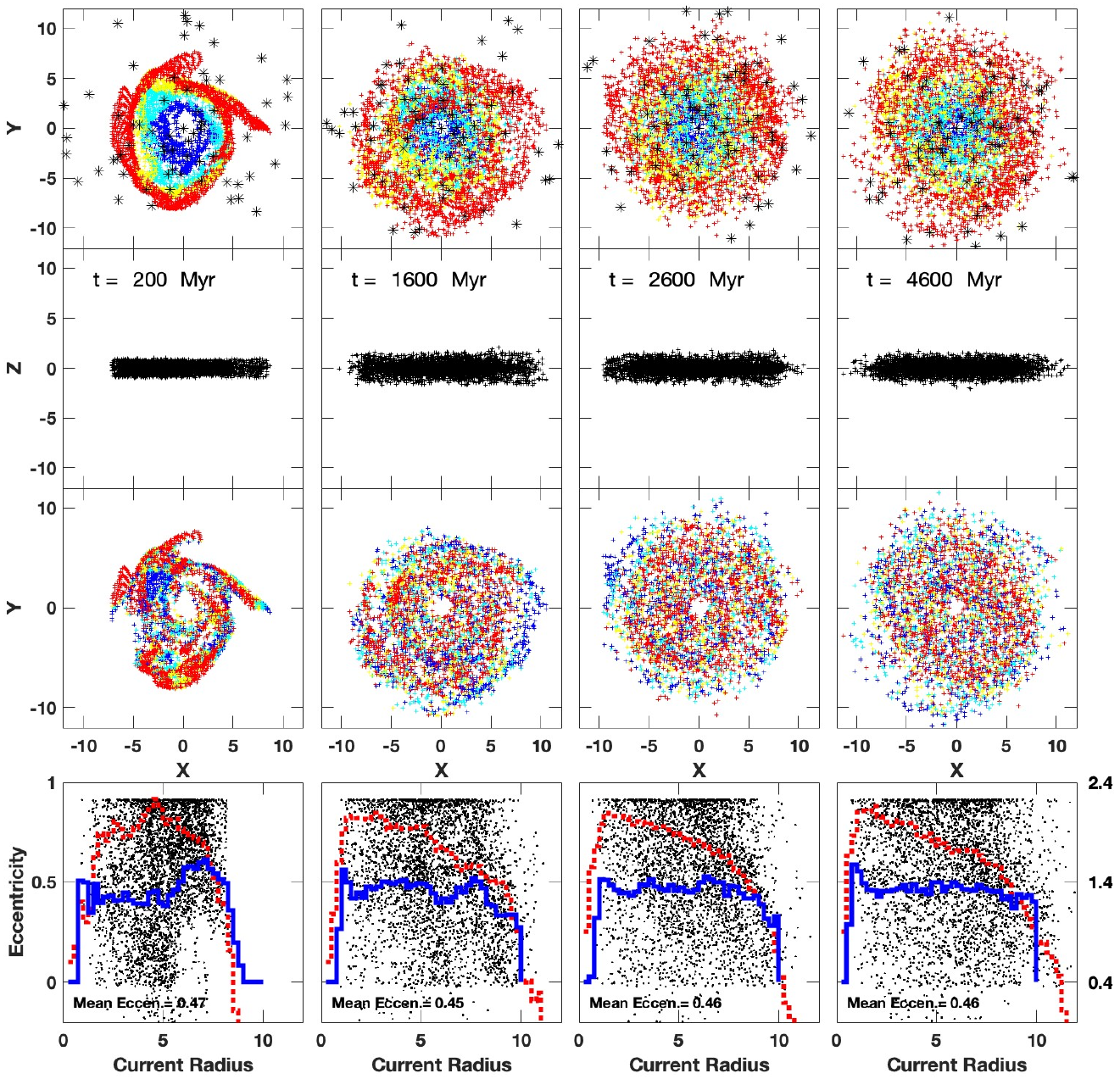}
\caption{Evolution of the fiducial (Normal) model. Each row shows four snapshots at times (in typical disk units described above) shown in the second row. The top row shows x-y (disk plane) views of the star particles and the massive clumps (asterisks). The star particles are color coded and binned by their initial radii, and only half of those used in the model are plotted. Extensive mixing is evident at late times. The second row shows x-z views of the disk, showing the effects of vertical scattering. The third row shows the disk plane again, but with color coding of the eccentricity; blue colors are the lowest eccentricity, while red are the highest (blue for $e<0.3$, cyan for $0.3<e<0.5$, yellow for $0.5<e<0.65$, and red for $e>0.65$.) The fourth row shows two quantities. The red dotted curves give the surface density profiles; refer to the base ten log scale on the far right panel of the plots, and which applies to all of the panels. The evolution from the initial flat profile to near exponential profiles is evident. The dots show the eccentricities of the star particles up to a maximum of 0.95. The eccentricity evolution is also shown by the blue, solid histograms of the mean eccentricity versus radius. The global mean eccentricity is given at the bottom.}
\label{fig:f1}
\end{figure*}

\begin{figure*}[ht!]
\plotone{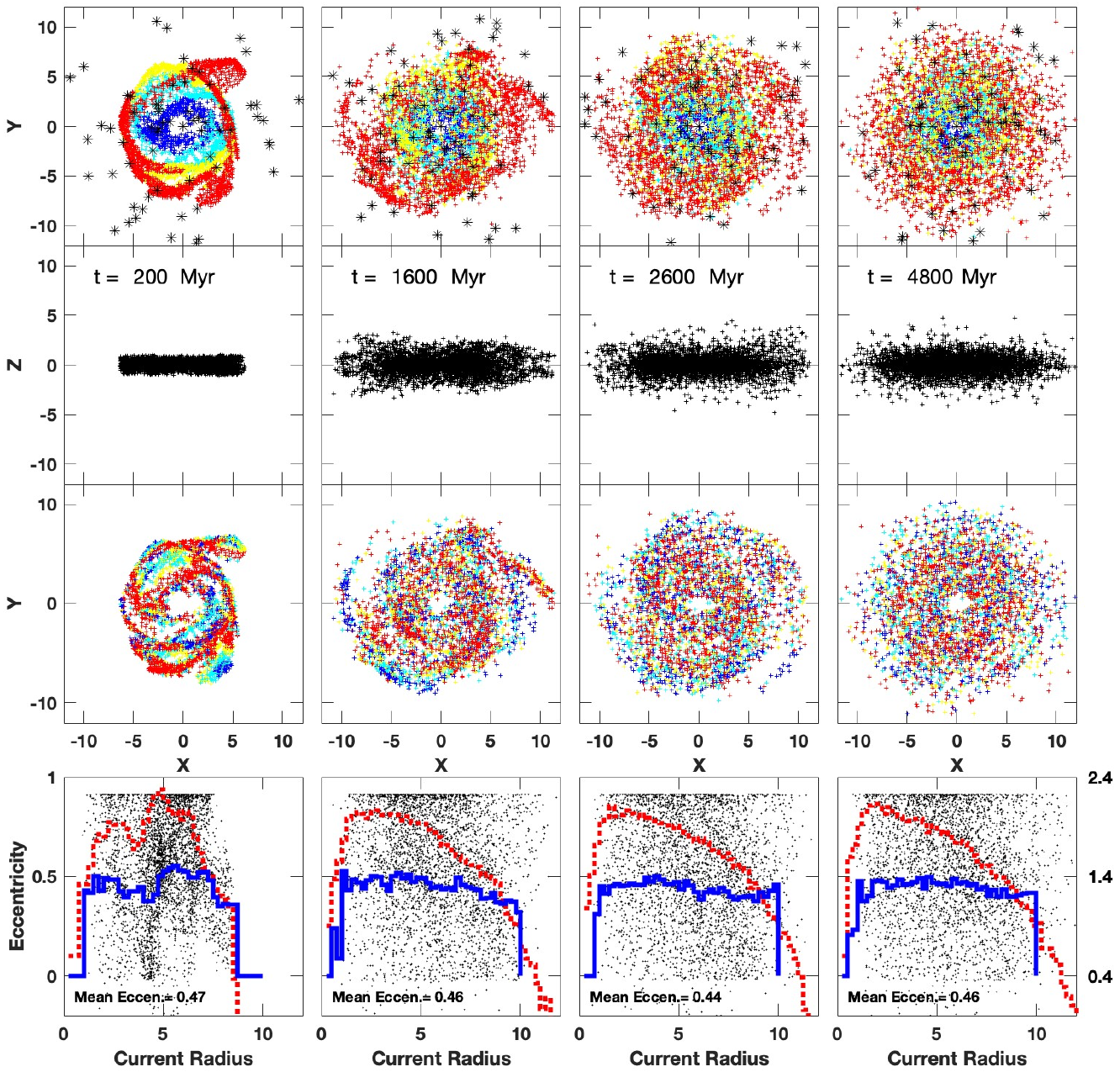}
\caption{Evolution in a (Anomalous) model that produced a fat disk. The model parameters and evolutionary graph sequence are the same as in Fig.~\ref{fig:f1}, but with slightly different stochastic initial conditions (see text). The result is similar surface profile and eccentricity distribution evolution, but much more vertical thickening. Many particles with high eccentricities leave the disk. A substantial population with low to moderate eccentricity remains at all radii in the disk, even at late times.}
\label{fig:f2}
\end{figure*}

In this section we consider the evolution of two realizations of a fiducial model of the scattering disk system. These models are essentially the same except for different values of the random variables in the initialization. (These initial random variables are: the initial particle azimuth, the initial particle vertical position as a fraction of the maximum of 0.3 units, and the initial particle vertical velocity as a fraction of the maximum of 1.5 units.) They will be compared to show the range of variation of the outcomes as a function of small initial setup changes.  We begin by considering the evolution of the first model as illustrated in Fig.~\ref{fig:f1}. For reasons that will be described, we will call this model run the Normal, and the second one (Fig.~\ref{fig:f2}) the Anomalous run. 

The rows of Fig.~\ref{fig:f1} show four snapshots in the evolution of various quantities. The first row shows the primary x-y plane of the disk with color-coded star points and asterisks marking the clump locations. Specifically, the four colors are derived by binning the initial star radii, and successive snapshots show the mixing resulting from the scattering. It is clear that the mixing of outer disk particles is extensive, but less for initial inner disk particles. The second row shows the vertical (x-z) growth of the disk. The vertical scattering is generally of second order relative to planar scattering, and generally modest, as in this run. 

The relatively mild vertical heating observed in many of our other simulations run over the same timescale suggests that cloud scattering is less likely to induce strong bulge components or thick disks compared to violent dynamical heating from mergers or bar buckling. This aligns with observational results showing thin exponential disks in low-mass galaxies with no classical bulge component \citep{kormendy16}.

The third row of Fig.~\ref{fig:f1} illustrates the development of orbital eccentricities across the disk. In this row eccentricities in the range (-0.05, 0.3) are colored blue, those in the range (0.3, 0.5) are colored cyan, those in the range (0.5, 0.65) are colored yellow, and those in the range (0.65, 0.99) are colored red. (Note: these are p-ellipse eccentricities, see \citealt{struck06}). The initial blue dominance of circular orbits evolves quickly to a broad distribution across the disk,as eccentricities initially increase rapidly, and eccentric particles spread across the disk. 

The fourth row of Fig.~\ref{fig:f1} extends our view of the eccentricity distribution by showing eccentricities versus radius for the stars (black points), and the evolution of the surface density profile (solid red curve). (The logarithmic scale (base ten) on the right-most panel applies to all of the surface density curves.) The global mean eccentricity of the stars is also given, and the blue, solid histograms show the mean eccentricity as a function of radius. Both the global value and the histogram bins are computed for all stars in the eccentricity range (0.0, 0.80). Stars with higher values of the eccentricity are viewed as essentially having escaped from the disk. There is also a maximum eccentricity of 0.95 allowed from the polynomial approximation described above. The surface density curves show that the exponential form develops promptly, initially as a Type II (double exponential) profile and later as more nearly a Type I (single exponential). The eccentricity distribution broadens quickly over the full range of values, and retains that broad distribution as the disk expands in radius at later times. In fact, at the start of a run a large number of particles are scattered up to eccentricities of more than 0.80, e.g., more than 50\% at the largest clump masses. However, as we will see in more detail below, the large eccentricity fraction and the eccentricity distribution reach a near-steady state quite quickly. That is, the ongoing scattering process does not drive the overall stellar ensemble to ever more eccentric orbits. 

Figure~\ref{fig:f2} shows a similar simulation, the Anomalous run, with slightly different initial conditions. However, there are substantial differences between the two, especially in the disk thicknesses by the end of the runs. First a thickening disk develops and by the end of the run some stars have been scattered into the halo. This run is called Anomalous because this behavior is unusual among these runs. 

The differences are unlikely the result of divergent evolution due to star scattering off single clumps. As evident in the models of \citet{struck17} and \citet{wu20}, clump scattering is amplified by chance alignments of multi-clump combinations. These alignments can produce massive local concentrations or transient spiral segments. These concentrations are more persistent and have greater effect in the Anomalous run of Figure~\ref{fig:f2}. Their appearance depends sensitively on initial conditions, so small random differences in these can generate large-scale differences in late time disk structure. The cases shown in Figs.~\ref{fig:f1} and \ref{fig:f2} are extremes among our runs, with a ratio of mean vertical (z) values of 0.63/0.39 (or 1.3 kpc and 0.78 kpc in the disk galaxy units above) at the end of the runs. The effect does not depend sensitively on clump mass, but smaller clump mass runs need to involve more clumps in the large scale asymmetry.  

This bifurcation between a minority of models that develop strong clump alignments and subsequent disk heating, versus those that do not, does not seem to be an obvious peculiarity of the simple model used here. However, it requires study in full N-body models, where it may be moderated by more complex (e.g., self-gravitational) processes. It is potentially an important phenomena, possibly driving a divergence between early and late-type disk galaxies. 

\subsection{Trajectories and Recircularization}

\begin{figure}[ht!]
\plotone{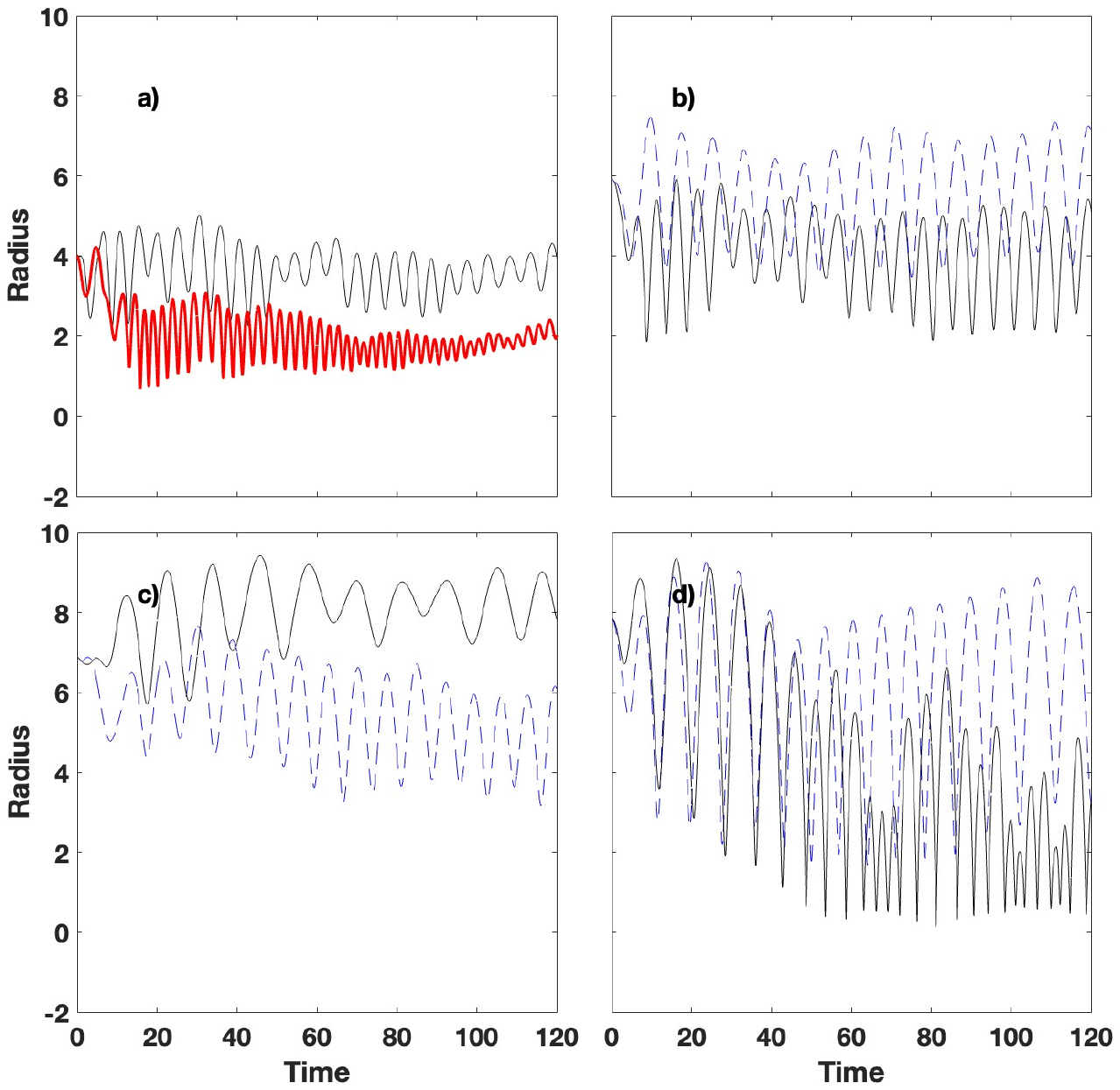}
\caption{Sample particle trajectories in dimensionless units. Each panel shows two trajectories beginning from very similar initial conditions. In Panel a) the two trajectories both show episodes of recircularization. The lower, bold, red one is an example of one that after several scattering and recircularization events ends in a nearly circular orbit at close to half its initial radius. In Panel b) the trajectories become moderately eccentric and remain so throughout the run. In Panel c) one trajectory evolves to significant eccentricity, while the second partially recircularizes at a somewhat larger than initial radius. In Panel d) one trajectory becomes highly eccentric, while the second partially recircularizes sporadically at a smaller than initial radius.}
\label{fig:f3}
\end{figure}

We noted above that our model disks tend toward a near steady state large-scale structural and an eccentricity distribution. This requires that scattering achieves a balance between events that increase star eccentricity and those that reduce it. We call the latter recircularization events. To better understand this phenomenon, we examine individual star radial trajectories in the fiducial run in Fig.~\ref{fig:f3}. The figure shows eight such trajectories, two in each of four panels, which illustrate the orbital evolution. The basic features of these orbits are described in the caption, here we note some general points. 

The first of these is that the majority of the trajectories change quite frequently, though stable orbits are not rare, e.g., see the black solid curve of panel c). Most of these changes are small, single clump scatterings. However, the curves of panel d) show a large jump in eccentricity at early times. The black, solid curve in that panel shows an example of recircularization at a considerably smaller guiding center radius than the initial, like the bold, red one in panel a). The black curve in panel c) shows a modest recircularization around a larger than initial guiding center. This same general behavior in individual trajectories is also seen in the N-body models of \citet{wu20}. 

Finally, we note that while these particular trajectories were chosen to illustrate certain characteristics, none of them are especially rare. A visual inspection of approximately 100 trajectories selected from a dozen runs revealed that about 20\% experienced a recircularization event of large magnitude over the course of the run (120 time units). Here, large magnitude is roughly $> 10\%$, with typical values of around $20\%$, with a very broad distribution. However, as evident in Fig.~\ref{fig:f3}, most trajectories are quite stable for large portions of the runs, though as noted, small increases or decreases of eccentricity are common.

\section{Systematics}

In the previous section we described the general characteristics of a fiducial scattering model, and something of the range of evolution by comparing to a second realization. In this section we consider an ensemble including several additional models to get a better understanding of the range of steady structures to which the scattering disk evolves, as well as the role of recircularization.

\begin{figure}
\plotone{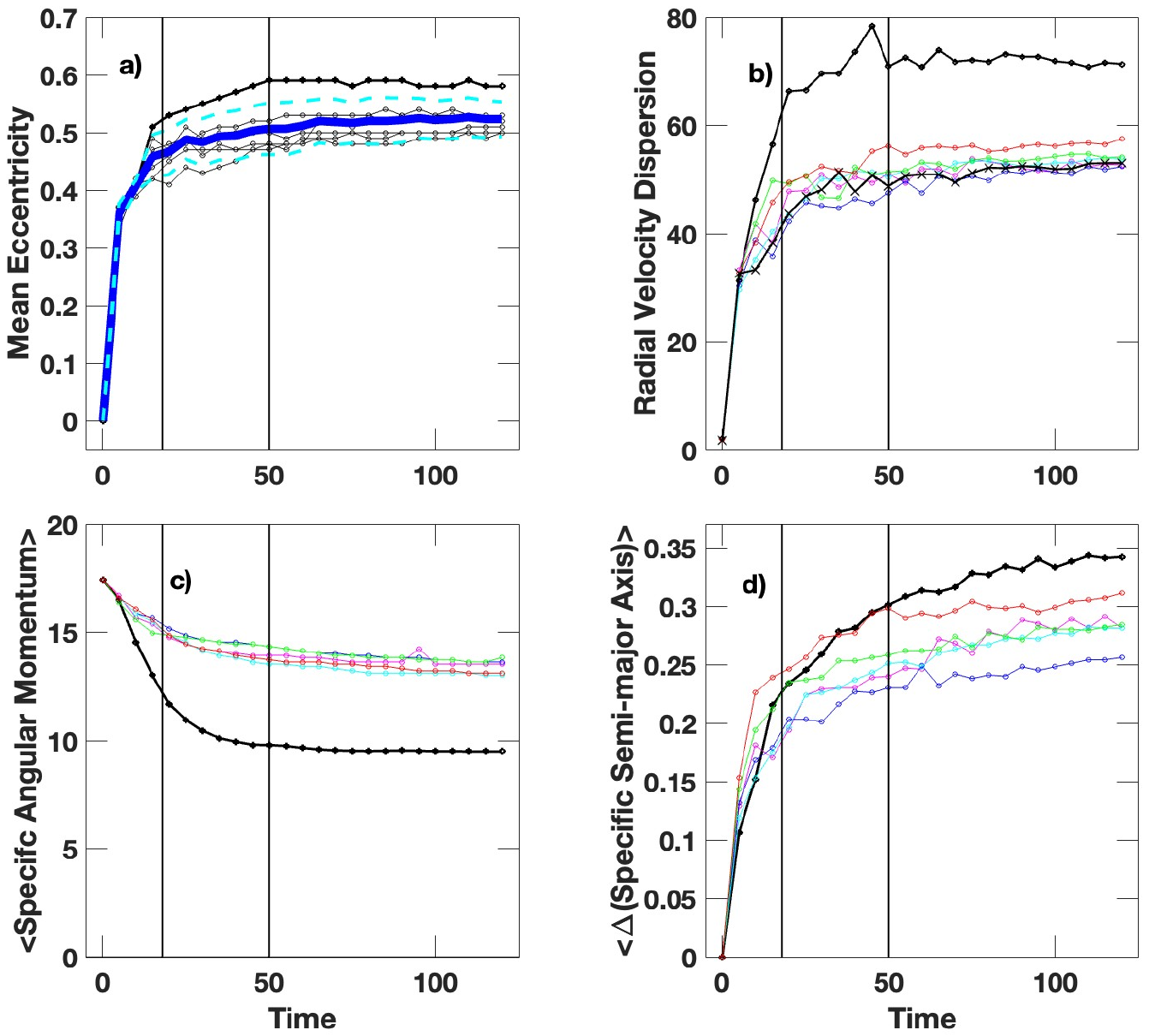}
\caption{Evolution of various quantities describing the orbit structure in six model discs, all with individual clump masses of 0.010 units. Time is in dimensionless units, except for panel b), multiply by 40 Myr for the typical disk time units above. Panel a) The evolution of mean orbital eccentricity for selected particles in 6 different runs of the simulation (see text for particle-inclusion criterion). Thin black curves show the results of the 6 runs. The thick blue line shows the mean, and the dashed cyan lines show one standard deviation from the mean. The vertical lines show estimates of the rapid increase timescale of the eccentricity (leftmost) and the onset of the steady state of the eccentricity and its mean (rightmost). These lines are replicated in the other panels. We note also that at a representative radius of 4.0 units the circular orbital period is about 25 units or about 220 Myr in the typical disk scaling, and roughly scales with radius. Panel b) The mean specific radial velocity dispersion for all particles in the model runs (with arbitrary colors). This panel contains an additional black curve with points marked by x symbols, showing the averaged 10times model. This model has clump masses reduced by a factor of 10 ($m_{cl} = 0.001$ units), and the number of clumps increased by the same factor. The number of stars is reduced in each 10times model run, so the curve here represents the average of four such runs.  The vertical scale corresponds to the example units for disk galaxies in km s$^{-1}$ from Sec. 2.2 above. The velocity dispersion scale for the dwarf galaxies example (not shown) is nearly the same. Panel c) Specific angular momentum evolution for each run averaged over all particles. Panel d) Average relative change as a function of time in the semi-major axes of all particles. This quantity is the average of the ratios of the difference between the semi-major axis of all particles at the time plotted and its semi-major axis near the beginning of the run, at $t = 0.1$ units, divided by the latter quantity. This is equivalent to the mean relative gyro-center evolution of the particles.}
\label{fig:f4}
\end{figure}

Panel a) of Fig.~\ref{fig:f4} shows the mean eccentricity evolution of six model runs (thin curves), the mean eccentricity evolution of the ensemble (thick blue curve) and the dispersion around that (dashed cyan curves). The mean eccentricities are computed as described in Sec. 3.1 in connection with Fig.~\ref{fig:f1}. The eccentricities rise rapidly in all models as clump scattering begins, on a timescale estimated in the figure followed by a period of slower increase (also marked in the figure), often with small oscillations. Then the eccentricities generally settle to the near steady state, with little dispersion, except for one case corresponding to Fig.~\ref{fig:f2}, highlighted in each panel by a thicker curve, and discussed further below. 

Panel b) provides another complementary view of this relaxation process. It shows the evolution of the radial velocity dispersion averaged over all of the particles. As in panel a) there is a rapid rise initially before the near steady state is reached. The anomalous run of panel a) is even further from the other runs in this plot.

An additional black solid curve, with data points marked by `x's, is shown in panel b). This represents the results of models with the clump mass reduced by a factor of 10 ($m_{cl} = 0.001$ units), and the number of clumps increased by the same factor. To maintain a reasonable computation time, the number of stars is also reduced by a factor of 10. On the other hand, to maintain statistical accuracy, multiple runs with the same clump initial conditions are averaged for analysis. We will refer to this as the `10times' model. The black solid curve of panel b) is an average of four runs. It is interesting that this averaged model evolves similarly to the fiducial models in both this and the other panels, though the 10times curve is not shown in these panels. The net scattering and its effects are comparable in all cases. Even aside from the Anomalous case, the final mean velocity dispersions are rather high. However, these averages include stars that have been scattered into orbits that are more characteristic of thick disks or a stellar halo. Nonetheless, the final row in Figs.~\ref{fig:f1}, \ref{fig:f2} show that in the near steady state a modest, old low eccentricity disk population remains.

Panel c) of the figure shows the evolution of the total angular momentum divided by the particle number for the ensemble runs. The run represented by the lowest curve is again the Anomalous run shown in Fig.~\ref{fig:f2}. Angular momentum is not conserved in these models because the scattering centers are on fixed circular orbits, and so, can remove angular momentum from the test particles with no effect on their own motion. Panel c) shows that the effect is largest at the beginning, settling phase, of the runs and small at late times. 

Examination of Fig.~\ref{fig:f2} and other snapshots reveals the cause of the anomalous curve - most of the clumps are broadly aligned in a rather loose oval form for an extended time. This global asymmetry of the clumps on fixed orbits subtracts angular momentum from the stars. We view it as analogous to a passing satellite. The highest curve in panel c) derives from the run shown in Fig.~\ref{fig:f2}. Many of the runs develop transient spiral arms or clump complexes the latter of which can affect the total angular momentum to some degree. The Anomalous run develops no such irregularity, and loses less angular momentum than the others. Evidently, the stochasticity of clump alignments has a significant effect on disk heating and thickness, though not on the mean surface density evolution. The convergence of mean eccentricity evolutions means that there is little effect on recircularization. 

Panel d) shows the relative change in semi-major axis ($(a - a_o)/(a_o)$) since a time at the beginning of the run (i.e., at $t_o = 0.1$ units and $a_o = a(t_o)$), and averaged over all of the particles with semi-major axis less than 10.0. The model evolutions of this quantity are similar to those of the previous panels, i.e., a fast rise, followed by an approach to a near steady state. The panel does show that on a percentage basis the mean change of particle gyro-center is less than that of the mean eccentricity. 

\begin{figure}
\plotone{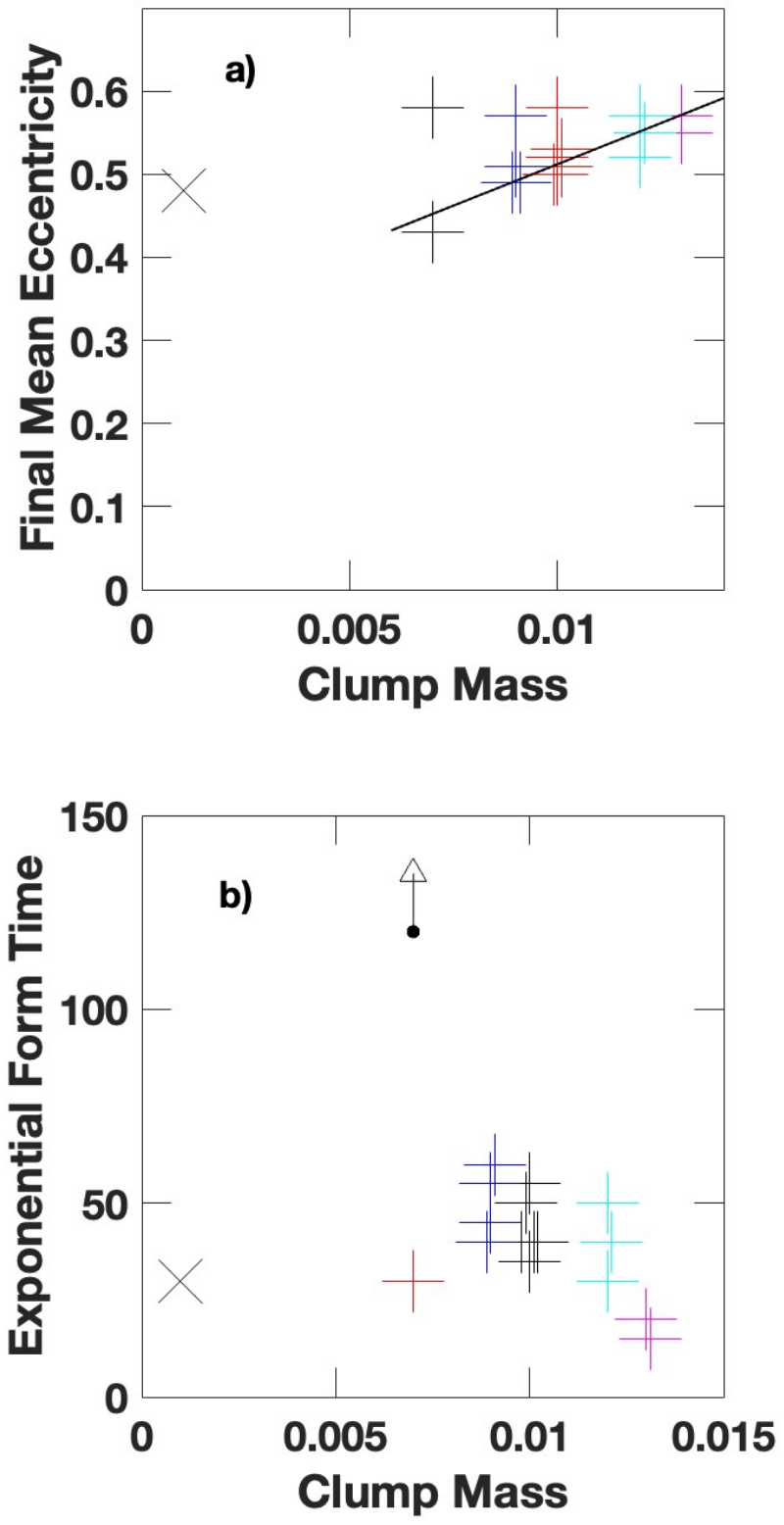}
\caption{Panel a) shows end-of-run eccentricities for models with a range of clump masses in dimensionless units, but constant clump number. The solid line shows a least squares fit to all of the models except that represented by the upper leftmost cross (+) and the averaged 10times model (x). Panel b) shows estimates of the time to form an exponential profile (in dimensionless units multiply by 40 Myr for typical disk units above) in runs with a range of clump masses. Crosses show estimates estimated visually from run outputs, while the arrow shows the end-of-run time for a runs that did not completely establish an exponential profile. The criterion for establishing an exponential was a straight line running over a factor of about 3 in radius in a log(surface density) versus radius plot, and a clear change in surface density within that radial range. The point in the lower left corresponds to that in the upper left of panel a) and the averaged 10times model is shown by the x symbol.}
\label{fig:f5}
\end{figure}

All of the above results have been derived from runs with a single clump mass (except for the 10times run), raising the possibility that the evolution to a near steady state and the recircularization process is not generic. Figure~\ref{fig:f5}, panel a) shows the end-of-run mean eccentricities of 19 models spanning a range of masses. We find that not only do all of these models evolve to a near steady state, but as the figure shows the mean eccentricity of that state increases only slowly with clump mass; a least squares linear fit to all the models, except the anomalous one in the upper left and the 10times model, is given by the solid line. The figure further shows modest dispersion around that mean, except at the lowest masses. 

Panel b) of Fig.~\ref{fig:f5} shows that the timescale for reaching near equilibrium decreases slowly with clump mass (for fixed clump number), except at the high and low mass ends. This timescale is estimated as the time when a given run first develops a straight line segment extending over a factor of a few in radius in the log(surface density) versus radius plot, and with a significant density change.  The timescale falls off nonlinearly at high masses. In the latter case, the clump masses are large enough to violate some of the underlying assumptions of the model, and so are probably not realistic. In the both cases, there is a large stochastic factor in this time scale, apparently another consequence of the large role that clump agglomerations play in the cumulative scattering process. The main conclusion from this figure is that the model disk evolution with recircularization to a near steady state is generic.

\section{Cloud scattering versus Spiral Arm Scattering}

The cloud-star scattering experiments discussed in the previous section changed an initially flat stellar profile into a tapered profile that approached an exponential after a sufficiently long time. The velocity dispersion was higher than in real galaxies because the clouds were relatively massive and each one strongly scattered its nearby stars.

Churning at the corotation resonance of a spiral arm can also move stars radially, and the velocity dispersion does not increase as much \citep{sellwood02}, suggesting that the churning model is more favorable. For churning, a star that approaches corotation from a smaller radius is attracted outward by the spiral, and then after reaching a larger radius, moves more slowly than the spiral and falls back in azimuth. In the corotating frame, the star makes a horseshoe-shape orbit from prograde inside to retrograde outside corotation. 

We showed in \cite{wu22} that the same horseshoe orbits occur for weak interactions between a cloud and a star, with the cloud taking the place of the corotation point in the spiral arm.  A problem with cloud scattering is that the scattering can be much stronger than spiral arm corotation scattering if the cloud is massive.  This is likely to be the case for young galaxies with giant clumps of star formation \citep{bournaud07}. Then the stellar velocity dispersion can get large, and the result is more like random scattering than churning. This problem may not occur for lower mass clouds that make up for their weak scatterings by an increased number.  The current paper attempts to move in this direction, but the clouds are not low enough in mass or numerous enough yet to keep the stellar dispersion low while still migrating stars outward. 

There are several difficulties with spiral-arm scattering as an explanation for exponential radial profiles. First, not all exponential disks have spiral arms. Dwarf irregular galaxies and the outer parts of spiral galaxies are examples (Sect. I). Second, even with spiral arms, the corotation resonance occupies only a small range of radii, close to where the Toomre Q parameter is low and the disk is most unstable. The exponential can extend much further than this, sometimes up to 10 scale lengths (Sect. I). 

A third possible problem is that although spirals may churn stars near corotation without much increase in velocity dispersion, they still increase the stellar dispersion at their Lindblad resonances. \cite{hamilton2024} note, following \cite{frankel20}, that the ratio of the radial action change to the angular momentum change in the Solar neighborhood over the last 6 Gyr is small, $\sim0.1$. This implies that Milky Way spirals have to be unusually weak at their Lindblad resonances, i.e. they have to be highly concentrated at corotation or shorter than the usual resonance-to-resonance length of a spiral arm. \cite{zhang25} also note that orbital migration is much stronger than heating, implying highly asymmetric scattering. \cite{mccluskey25} conclude the same thing for the Milky Way, which has an unusually low ratio of velocity dispersion to orbit speed among local galaxies. 

For the spiral arm scattering model, \cite{hamilton2024} state that the small ratio of heating to scattering requires a history of spirals that were more open, more concentrated around corotation, and/or shorter-lived than the current Milky Way spirals. The openness constraint allows the spirals to apply more azimuthal forcing to influence the angular momentum distribution without applying as much radial forcing, which goes into the velocity dispersion. The concentration around corotation removes the scattering at the Lindblad resonances, where the spiral-to-spiral period equals some resonant multiple of the epicyclic period and spirals pump epicyclic (random) motions directly. The short life applies to what \cite{hamilton2024} call the resonant regime, which is where the arm-to-arm time is less than the duration of the spirals, and this is less than half the libration time. The libration time was defined for horseshoe orbits and is the arm-to-arm time divided by the  ratio of the square root of the spiral potential to the orbit speed. 

All three of these constraints can be more easily satisfied for cloud-star scattering than spiral arm scattering. Consider the third constraint first. Clouds are much more numerous and much weaker than spiral arms. The ratio of the square root of the cloud potential energy to the orbit speed is $\sim0.01$, because the internal cloud velocity dispersion, in the case of virialized GMCs for example, is only a few km s$^{-1}$ or less, and the orbit speed is $\sim100$ times this. This lower scattering potential is offset, however, by an increased number of scatterers (clouds), which is like increasing the number of spiral arms, $m$, in equation (5) in \cite{hamilton2024}. These two changes keep the libration time about the same for the cloud model as the spiral model, but the higher effective $m$ decreases greatly the resonant time, which is the time between cloud scatterings in the cloud model.  Thus the third  solution suggested by \cite{hamilton2024} is plausible for the cloud model. A definitive assessment of the relative roles of cloud scattering and spiral-arm scattering will require fully self-consistent simulations that include both clouds and transient spiral structure.

Clouds also have all of their forcing at corotation and little forcing at their Lindblad resonances, because they are relatively small in radial extent (i.e., 10-100 pc) while their resonances are several kpc away. 
Clouds are also everywhere in a gas disk, which often extends far beyond the stellar disk for a spiral galaxy and dominates the stellar disk in a dwarf irregular. Thus, they are always available to scatter stars even when there are no spirals. 

\section{Conclusions}
In this paper we presented a set of numerical experiments showing how gravitational scattering by interstellar clouds leads to the formation of exponential stellar disks. Cloud-star scattering has long been known to heat a stellar disk, increasing the eccentricities of stars \citep{spitzer51,sun25}. It is one of several possible heating mechanisms \citep{garzon24}. Cloud scattering stalls once stars move off the thin cloud layer, so there is a natural regulation of the velocity dispersion \citep{wielen85,struck17}. Stars that remain in the gas layer continue to scatter, but there appears to be another process that limits even this heating. We show here that most stellar eccentricities decrease with time from numerous and continuous weak scatterings, while a small fraction of the stars increase their eccentricities when they interact more strongly with a cloud. The result is an equilibrium distribution function of stellar eccentricity. 
A novel insight of this work is the role of re-circularization in establishing a steady-state eccentricity distribution. Stars on more eccentric orbits tend to move above the gas layer, thereby reducing their interaction cross-section with interstellar clouds. This leads to a natural selection effect, where stars that remain near the midplane are more likely to experience weak scatterings that reduce their eccentricities. This dynamical feedback resembles processes seen in equilibrium-regulated systems, such as diffusion-limited growth (\citealt{wielen85, struck17}). 

The simulations demonstrate that exponential profiles form quite rapidly and are maintained over time, even as stars migrate radially outward and the disk expands. While the vertical thickening varies between runs due to stochastic differences (e.g., cloud alignments), in most cases the vertical heating remains modest. This suggests that cloud scattering may not be as efficient in producing thick disks or bulges as mergers or bar buckling. The runs show a persistent population of stars with low-to-moderate eccentricities at all radii, and the surface density remains close to exponential.

We note, however, that some of these evolutionary details are the result of the idealized model and initial conditions. For instance the rapid rise in eccentricity must be due in large part to the perfectly circular initial orbits of the particles. An initial disk with large initial random velocities would likely react less strongly and more slowly to early clump scattering. Young disks are likely to have high dispersions \citep[e.g.,][]{donkelaar2022}. Scattering to the outer disk, in particular, may take longer. However, in such cases recircularization may begin earlier, and drive the disk to a similar equilibrium to that found above.

We find that the equilibrium eccentricity distribution is nearly independent of clump mass for constant clump mass, and the timescale to reach it does not depend sensitively on that factor. Therefore, the idealization of a single clump mass in the models may not be very important. The equilibrium eccentricity distribution is nearly flat at almost all times, except for small peaks at zero and at the maximum allowed eccentricity. The recircularization behavior is robust, occurring even in disks with different cloud masses and in the presence of different scattering histories. This is evident from both individual stellar trajectories and ensemble statistics, showing that decreases in eccentricity are common. 


The disk angular momentum tends to decrease slightly in these models, even as the average semi-major axes get larger due to the changing angular momentum. In a minority of runs temporary asymmetries develop in the clump distribution. These lead to more vigorous scattering, and if not too transient, to larger eccentricities and velocity distributions, and a bifurcation in the quantities describing the near steady states. Nonetheless, the dominant results are that steady eccentricity distributions and exponential profiles persist regardless.

These findings support the idea that cloud scattering can operate as a universal mechanism to build and maintain exponential disks, without requiring large-scale spiral arms or bars. The results are particularly relevant to dwarf and high-redshift galaxies where turbulence and clumpiness dominate the ISM. The model produces several features that can be compared to observations including eccentricity distributions, mixing behaviors, and exponential profile development timescales. The model provides a physical framework that may complement other processes like viscous evolution or star formation feedback.
\\

\begin{center} {\bf Acknowledgements} \end{center}

The Astrophysical Data Service (https://ui.adsabs.harvard.edu/classic-form/) and the arXiv preprint repository (https://arxiv.org/) were used throughout the preparation of this paper. We are grateful to an anonymous reviewer for helpful suggestions.

\bibliography{refs}{}

@ARTICLE{hamilton2024,
       author = {{Hamilton}, Chris and {Modak}, Shaunak and {Tremaine}, Scott},
        title = "{Why is the Galactic disk so cool?}",
      journal = {arXiv e-prints},
         year = 2024,
        month = nov,
          eid = {arXiv:2411.08944},
        pages = {arXiv:2411.08944},
          doi = {10.48550/arXiv.2411.08944},
archivePrefix = {arXiv},
       eprint = {2411.08944},
 primaryClass = {astro-ph.GA},
       adsurl = {https://ui.adsabs.harvard.edu/abs/2024arXiv241108944H}
}

@ARTICLE{donkelaar2022,
       author = {{van Donkelaar}, Floor and {Agertz}, Oscar and {Renaud}, Florent},
        title = "{From giant clumps to clouds - II. The emergence of thick disc kinematics from the conditions of star formation in high redshift gas rich galaxies}",
      journal = {\mnras},
         year = 2022,
        month = may,
       volume = {512},
       number = {3},
        pages = {3806-3814},
          doi = {10.1093/mnras/stac692},
archivePrefix = {arXiv},
       eprint = {2110.13165},
 primaryClass = {astro-ph.GA},
       adsurl = {https://ui.adsabs.harvard.edu/abs/2022MNRAS.512.3806V}
}

@ARTICLE{salcedo2025,
       author = {{Espejo Salcedo}, J.~M. and {Pastras}, S. and {V{\'a}cha}, J. and {Pulsoni}, C. and {Genzel}, R. and {F{\"o}rster Schreiber}, N.~M. and {Jolly}, J.-B. and {Barfety}, C. and {Chen}, J. and {Tozzi}, G. and {Liu}, D. and {Lee}, L.~L. and {Wuyts}, S. and {Tacconi}, L.~J. and {Davies}, R. and {{\"U}bler}, H. and {Lutz}, D. and {Wisnioski}, E. and {Shangguan}, J. and {Lee}, M. and {Price}, S.~H. and {Eisenhauer}, F. and {Renzini}, A. and {Nestor Shachar}, A. and {Herrera-Camus}, R.},
        title = "{Galaxy morphologies at cosmic noon with JWST: A foundation for exploring gas transport with bars and spiral arms}",
      journal = {\aap},
         year = 2025,
        month = aug,
       volume = {700},
          eid = {A42},
        pages = {A42},
          doi = {10.1051/0004-6361/202554725},
archivePrefix = {arXiv},
       eprint = {2503.21738},
 primaryClass = {astro-ph.GA},
       adsurl = {https://ui.adsabs.harvard.edu/abs/2025A&A...700A..42E}
}

@ARTICLE{kuhn24,
       author = {{Kuhn}, Vicki and {Guo}, Yicheng and {Martin}, Alec and {Bayless}, Julianna and {Gates}, Ellie and {Puleo}, AJ},
        title = "{JWST Reveals a Surprisingly High Fraction of Galaxies Being Spiral-like at 0.5 {\ensuremath{\leq}} z {\ensuremath{\leq}} 4}",
      journal = {\apjl},
         year = 2024,
        month = jun,
       volume = {968},
       number = {2},
          eid = {L15},
        pages = {L15},
          doi = {10.3847/2041-8213/ad43eb},
archivePrefix = {arXiv},
       eprint = {2312.12389},
 primaryClass = {astro-ph.GA},
       adsurl = {https://ui.adsabs.harvard.edu/abs/2024ApJ...968L..15K}
}

@ARTICLE{struck15,
       author = {{Struck}, Curtis},
        title = "{Natural orbit approximations in single power-law potentials}",
      journal = {\mnras},
         year = 2015,
        month = jan,
       volume = {446},
       number = {3},
        pages = {3139-3149},
          doi = {10.1093/mnras/stu2342},
archivePrefix = {arXiv},
       eprint = {1411.1022},
 primaryClass = {astro-ph.GA},
       adsurl = {https://ui.adsabs.harvard.edu/abs/2015MNRAS.446.3139S}
}

@ARTICLE{Ono25,
       author = {{Ono}, Yoshiaki and {Ouchi}, Masami and {Harikane}, Yuichi and {Yajima}, Hidenobu and {Nakajima}, Kimihiko and {Fujimoto}, Seiji and {Nakane}, Minami and {Xu}, Yi},
        title = "{Morphological Demographics of Galaxies at z {\ensuremath{\sim}} 10{\textendash}16: Log-normal Size Distribution and Exponential Profiles Consistent with the Disk Formation Scenario}",
      journal = {\apj},
         year = 2025,
        month = oct,
       volume = {991},
       number = {2},
          eid = {222},
        pages = {222},
          doi = {10.3847/1538-4357/adfc4d},
       adsurl = {https://ui.adsabs.harvard.edu/abs/2025ApJ...991..222O}
}

@ARTICLE{donghia16,
       author = {{D'Onghia}, E. and {Madau}, P. and {Vera-Ciro}, C. and {Quillen}, A. and {Hernquist}, L.},
        title = "{Excitation of Coupled Stellar Motions in the Galactic Disk by Orbiting Satellites}",
      journal = {\apj},
         year = 2016,
        month = may,
       volume = {823},
       number = {1},
          eid = {4},
        pages = {4},
          doi = {10.3847/0004-637X/823/1/4},
archivePrefix = {arXiv},
       eprint = {1511.01503},
 primaryClass = {astro-ph.GA},
       adsurl = {https://ui.adsabs.harvard.edu/abs/2016ApJ...823....4D}
}

@ARTICLE{sok25,
       author = {{Sok}, Visal and {Muzzin}, Adam and {Tan}, Vivian Yun Yan and {Asada}, Yoshihisa and {Brada{\v{c}}}, Maru{\v{s}}a and {Martis}, Nicholas S. and {Noirot}, Ga{\"e}l and {Sarrouh}, Ghassan T.~E. and {Sawicki}, Marcin and {Willott}, Chris J. and {Withers}, Sunna and {Berek}, Samantha C. and {Myers}, Katherine},
        title = "{The Stellar Mass and Age Distributions of Star-Forming Clumps at $0.5 < z < 5$ in JWST CANUCS: Implications for Clump Formation and Destruction}",
      journal = {arXiv e-prints},
         year = 2025,
        month = sep,
          eid = {arXiv:2509.25363},
        pages = {arXiv:2509.25363},
archivePrefix = {arXiv},
       eprint = {2509.25363},
 primaryClass = {astro-ph.GA},
       adsurl = {https://ui.adsabs.harvard.edu/abs/2025arXiv250925363S}
}

@ARTICLE{kormendy16,
       author = {{Kormendy}, John and {Freeman}, K.~C.},
        title = "{Scaling Laws for Dark Matter Halos in Late-type and Dwarf Spheroidal Galaxies}",
      journal = {\apj},
         year = 2016,
        month = feb,
       volume = {817},
       number = {2},
          eid = {84},
        pages = {84},
          doi = {10.3847/0004-637X/817/2/84},
archivePrefix = {arXiv},
       eprint = {1411.2170},
 primaryClass = {astro-ph.GA},
       adsurl = {https://ui.adsabs.harvard.edu/abs/2016ApJ...817...84K}
}

@ARTICLE{freeman70,
       author = {{Freeman}, K.~C.},
        title = "{On the Disks of Spiral and S0 Galaxies}",
      journal = {\apj},
         year = 1970,
        month = jun,
       volume = {160},
        pages = {811},
          doi = {10.1086/150474},
       adsurl = {https://ui.adsabs.harvard.edu/abs/1970ApJ...160..811F}
}

@ARTICLE{berrier15,
       author = {{Berrier}, Joel C. and {Sellwood}, J.~A.},
        title = "{Smoothing Rotation Curves and Mass Profiles}",
      journal = {\apj},
         year = 2015,
        month = feb,
       volume = {799},
       number = {2},
          eid = {213},
        pages = {213},
          doi = {10.1088/0004-637X/799/2/213},
archivePrefix = {arXiv},
       eprint = {1412.0979},
 primaryClass = {astro-ph.GA},
       adsurl = {https://ui.adsabs.harvard.edu/abs/2015ApJ...799..213B}
}

@ARTICLE{erwin08,
       author = {{Erwin}, Peter and {Pohlen}, Michael and {Beckman}, John E.},
        title = "{The Outer Disks of Early-Type Galaxies. I. Surface-Brightness Profiles of Barred Galaxies}",
      journal = {\aj},
         year = 2008,
        month = jan,
       volume = {135},
       number = {1},
        pages = {20-54},
          doi = {10.1088/0004-6256/135/1/20},
archivePrefix = {arXiv},
       eprint = {0709.3505},
 primaryClass = {astro-ph},
       adsurl = {https://ui.adsabs.harvard.edu/abs/2008AJ....135...20E}
}

@ARTICLE{herrmann13,
       author = {{Herrmann}, Kimberly A. and {Hunter}, Deidre A. and {Elmegreen}, Bruce G.},
        title = "{Surface Brightness Profiles of Dwarf Galaxies. I. Profiles and Statistics}",
      journal = {\aj},
         year = 2013,
        month = nov,
       volume = {146},
       number = {5},
          eid = {104},
        pages = {104},
          doi = {10.1088/0004-6256/146/5/104},
archivePrefix = {arXiv},
       eprint = {1309.0004},
 primaryClass = {astro-ph.CO},
       adsurl = {https://ui.adsabs.harvard.edu/abs/2013AJ....146..104H}
}

@ARTICLE{foyle08,
       author = {{Foyle}, Kelly and {Courteau}, St{\'e}phane and {Thacker}, Robert J.},
        title = "{An N-body/SPH study of isolated galaxy mass density profiles}",
      journal = {\mnras},
         year = 2008,
        month = jun,
       volume = {386},
       number = {4},
        pages = {1821-1844},
          doi = {10.1111/j.1365-2966.2008.13201.x},
archivePrefix = {arXiv},
       eprint = {0803.2716},
 primaryClass = {astro-ph},
       adsurl = {https://ui.adsabs.harvard.edu/abs/2008MNRAS.386.1821F}
}

@ARTICLE{wang22,
       author = {{Wang}, Enci and {Lilly}, Simon J.},
        title = "{The Origin of Exponential Star-forming Disks}",
      journal = {\apj},
         year = 2022,
        month = mar,
       volume = {927},
       number = {2},
          eid = {217},
        pages = {217},
          doi = {10.3847/1538-4357/ac49ed},
archivePrefix = {arXiv},
       eprint = {2201.04148},
 primaryClass = {astro-ph.GA},
       adsurl = {https://ui.adsabs.harvard.edu/abs/2022ApJ...927..217W}
}

@ARTICLE{hunter06,
       author = {{Hunter}, Deidre A. and {Elmegreen}, Bruce G.},
        title = "{Broadband Imaging of a Large Sample of Irregular Galaxies}",
      journal = {\apjs},
         year = 2006,
        month = jan,
       volume = {162},
       number = {1},
        pages = {49-79},
          doi = {10.1086/498096},
archivePrefix = {arXiv},
       eprint = {astro-ph/0509189},
 primaryClass = {astro-ph},
       adsurl = {https://ui.adsabs.harvard.edu/abs/2006ApJS..162...49H}
}

@ARTICLE{elmegreen12,
       author = {{Elmegreen}, Debra Meloy and {Elmegreen}, Bruce G. and {S{\'a}nchez Almeida}, Jorge and {Mu{\~n}oz-Tu{\~n}{\'o}n}, Casiana and {Putko}, Joseph and {Dewberry}, Janosz},
        title = "{Local Tadpole Galaxies}",
      journal = {\apj},
         year = 2012,
        month = may,
       volume = {750},
       number = {2},
          eid = {95},
        pages = {95},
          doi = {10.1088/0004-637X/750/2/95},
archivePrefix = {arXiv},
       eprint = {1203.2486},
 primaryClass = {astro-ph.CO},
       adsurl = {https://ui.adsabs.harvard.edu/abs/2012ApJ...750...95E}
}

@ARTICLE{morales11,
       author = {{Morales-Luis}, A.~B. and {S{\'a}nchez Almeida}, J. and {Aguerri}, J.~A.~L. and {Mu{\~n}oz-Tu{\~n}{\'o}n}, C.},
        title = "{Systematic Search for Extremely Metal-poor Galaxies in the Sloan Digital Sky Survey}",
      journal = {\apj},
         year = 2011,
        month = dec,
       volume = {743},
       number = {1},
          eid = {77},
        pages = {77},
          doi = {10.1088/0004-637X/743/1/77},
archivePrefix = {arXiv},
       eprint = {1109.0235},
 primaryClass = {astro-ph.CO},
       adsurl = {https://ui.adsabs.harvard.edu/abs/2011ApJ...743...77M}
}

@ARTICLE{sanchez13,
       author = {{S{\'a}nchez Almeida}, J. and {Mu{\~n}oz-Tu{\~n}{\'o}n}, C. and {Elmegreen}, D.~M. and {Elmegreen}, B.~G. and {M{\'e}ndez-Abreu}, J.},
        title = "{Local Tadpole Galaxies: Dynamics and Metallicity}",
      journal = {\apj},
         year = 2013,
        month = apr,
       volume = {767},
       number = {1},
          eid = {74},
        pages = {74},
          doi = {10.1088/0004-637X/767/1/74},
archivePrefix = {arXiv},
       eprint = {1302.4352},
 primaryClass = {astro-ph.CO},
       adsurl = {https://ui.adsabs.harvard.edu/abs/2013ApJ...767...74S}
}

@ARTICLE{filho13,
       author = {{Filho}, M.~E. and {Winkel}, B. and {S{\'a}nchez Almeida}, J. and {Aguerri}, J.~A. and {Amor{\'\i}n}, R. and {Ascasibar}, Y. and {Elmegreen}, B.~G. and {Elmegreen}, D.~M. and {Gomes}, J.~M. and {Humphrey}, A. and {Lagos}, P. and {Morales-Luis}, A.~B. and {Mu{\~n}oz-Tu{\~n}{\'o}n}, C. and {Papaderos}, P. and {V{\'\i}lchez}, J.~M.},
        title = "{Extremely metal-poor galaxies: The H I content}",
      journal = {\aap},
         year = 2013,
        month = oct,
       volume = {558},
          eid = {A18},
        pages = {A18},
          doi = {10.1051/0004-6361/201322098},
archivePrefix = {arXiv},
       eprint = {1307.4899},
 primaryClass = {astro-ph.CO},
       adsurl = {https://ui.adsabs.harvard.edu/abs/2013A&A...558A..18F}
}

@ARTICLE{pohlen06,
       author = {{Pohlen}, M. and {Trujillo}, I.},
        title = "{The structure of galactic disks. Studying late-type spiral galaxies using SDSS}",
      journal = {\aap},
         year = 2006,
        month = aug,
       volume = {454},
       number = {3},
        pages = {759-772},
          doi = {10.1051/0004-6361:20064883},
archivePrefix = {arXiv},
       eprint = {astro-ph/0603682},
 primaryClass = {astro-ph},
       adsurl = {https://ui.adsabs.harvard.edu/abs/2006A&A...454..759P}
}

@ARTICLE{hunt15,
       author = {{Hunt}, L.~K. and {Draine}, B.~T. and {Bianchi}, S. and {Gordon}, K.~D. and {Aniano}, G. and {Calzetti}, D. and {Dale}, D.~A. and {Helou}, G. and {Hinz}, J.~L. and {Kennicutt}, R.~C. and {Roussel}, H. and {Wilson}, C.~D. and {Bolatto}, A. and {Boquien}, M. and {Croxall}, K.~V. and {Galametz}, M. and {Gil de Paz}, A. and {Koda}, J. and {Mu{\~n}oz-Mateos}, J.~C. and {Sandstrom}, K.~M. and {Sauvage}, M. and {Vigroux}, L. and {Zibetti}, S.},
        title = "{Cool dust heating and temperature mixing in nearby star-forming galaxies}",
      journal = {\aap},
         year = 2015,
        month = apr,
       volume = {576},
          eid = {A33},
        pages = {A33},
          doi = {10.1051/0004-6361/201424734},
archivePrefix = {arXiv},
       eprint = {1409.5916},
 primaryClass = {astro-ph.GA},
       adsurl = {https://ui.adsabs.harvard.edu/abs/2015A&A...576A..33H}
}

@ARTICLE{munoz09,
       author = {{Mu{\~n}oz-Mateos}, J.~C. and {Gil de Paz}, A. and {Boissier}, S. and {Zamorano}, J. and {Dale}, D.~A. and {P{\'e}rez-Gonz{\'a}lez}, P.~G. and {Gallego}, J. and {Madore}, B.~F. and {Bendo}, G. and {Thornley}, M.~D. and {Draine}, B.~T. and {Boselli}, A. and {Buat}, V. and {Calzetti}, D. and {Moustakas}, J. and {Kennicutt}, Jr., R.~C.},
        title = "{Radial Distribution of Stars, Gas, and Dust in Sings Galaxies. II. Derived Dust Properties}",
      journal = {\apj},
         year = 2009,
        month = aug,
       volume = {701},
       number = {2},
        pages = {1965-1991},
          doi = {10.1088/0004-637X/701/2/1965},
archivePrefix = {arXiv},
       eprint = {0909.2658},
 primaryClass = {astro-ph.CO},
       adsurl = {https://ui.adsabs.harvard.edu/abs/2009ApJ...701.1965M}
}

@ARTICLE{bland05,
       author = {{Bland-Hawthorn}, J. and {Vlaji{\'c}}, M. and {Freeman}, K.~C. and {Draine}, B.~T.},
        title = "{NGC 300: An Extremely Faint, Outer Stellar Disk Observed to 10 Scale Lengths}",
      journal = {\apj},
         year = 2005,
        month = aug,
       volume = {629},
       number = {1},
        pages = {239-249},
          doi = {10.1086/430512},
archivePrefix = {arXiv},
       eprint = {astro-ph/0503488},
 primaryClass = {astro-ph},
       adsurl = {https://ui.adsabs.harvard.edu/abs/2005ApJ...629..239B}
}

@ARTICLE{wang16,
       author = {{Wang}, Jing and {Koribalski}, B{\"a}rbel S. and {Serra}, Paolo and {van der Hulst}, Thijs and {Roychowdhury}, Sambit and {Kamphuis}, Peter and {Chengalur}, Jayaram N.},
        title = "{New lessons from the H I size-mass relation of galaxies}",
      journal = {\mnras},
         year = 2016,
        month = aug,
       volume = {460},
       number = {2},
        pages = {2143-2151},
          doi = {10.1093/mnras/stw1099},
archivePrefix = {arXiv},
       eprint = {1605.01489},
 primaryClass = {astro-ph.GA},
       adsurl = {https://ui.adsabs.harvard.edu/abs/2016MNRAS.460.2143W}
}

@ARTICLE{bigiel12,
       author = {{Bigiel}, F. and {Blitz}, L.},
        title = "{A Universal Neutral Gas Profile for nearby Disk Galaxies}",
      journal = {\apj},
         year = 2012,
        month = sep,
       volume = {756},
       number = {2},
          eid = {183},
        pages = {183},
          doi = {10.1088/0004-637X/756/2/183},
archivePrefix = {arXiv},
       eprint = {1208.1505},
 primaryClass = {astro-ph.CO},
       adsurl = {https://ui.adsabs.harvard.edu/abs/2012ApJ...756..183B}
}

@ARTICLE{dutton09,
       author = {{Dutton}, Aaron A.},
        title = "{On the origin of exponential galaxy discs}",
      journal = {\mnras},
         year = 2009,
        month = jun,
       volume = {396},
       number = {1},
        pages = {121-140},
          doi = {10.1111/j.1365-2966.2009.14741.x},
archivePrefix = {arXiv},
       eprint = {0810.5164},
 primaryClass = {astro-ph},
       adsurl = {https://ui.adsabs.harvard.edu/abs/2009MNRAS.396..121D}
}

@ARTICLE{meert15,
       author = {{Meert}, Alan and {Vikram}, Vinu and {Bernardi}, Mariangela},
        title = "{A catalogue of 2D photometric decompositions in the SDSS-DR7 spectroscopic main galaxy sample: preferred models and systematics}",
      journal = {\mnras},
         year = 2015,
        month = feb,
       volume = {446},
       number = {4},
        pages = {3943-3974},
          doi = {10.1093/mnras/stu2333},
archivePrefix = {arXiv},
       eprint = {1406.4179},
 primaryClass = {astro-ph.GA},
       adsurl = {https://ui.adsabs.harvard.edu/abs/2015MNRAS.446.3943M}
}

@ARTICLE{weiner01,
       author = {{Weiner}, Benjamin J. and {Williams}, T.~B. and {van Gorkom}, J.~H. and {Sellwood}, J.~A.},
        title = "{The Disk and Dark Halo Mass of the Barred Galaxy NGC 4123. I. Observations}",
      journal = {\apj},
         year = 2001,
        month = jan,
       volume = {546},
       number = {2},
        pages = {916-930},
          doi = {10.1086/318288},
archivePrefix = {arXiv},
       eprint = {astro-ph/0008204},
 primaryClass = {astro-ph},
       adsurl = {https://ui.adsabs.harvard.edu/abs/2001ApJ...546..916W}
}

@ARTICLE{hunter11,
       author = {{Hunter}, Deidre A. and {Elmegreen}, Bruce G. and {Oh}, Se-Heon and {Anderson}, Ed and {Nordgren}, Tyler E. and {Massey}, Philip and {Wilsey}, Nick and {Riabokin}, Malanka},
        title = "{The Outer Disks of Dwarf Irregular Galaxies}",
      journal = {\aj},
         year = 2011,
        month = oct,
       volume = {142},
       number = {4},
          eid = {121},
        pages = {121},
          doi = {10.1088/0004-6256/142/4/121},
archivePrefix = {arXiv},
       eprint = {1107.5587},
 primaryClass = {astro-ph.CO},
       adsurl = {https://ui.adsabs.harvard.edu/abs/2011AJ....142..121H}
}

@ARTICLE{hillis16,
       author = {{Hillis}, Tristan J. and {Williams}, Benjamin F. and {Dolphin}, Andrew E. and {Dalcanton}, Julianne J. and {Skillman}, Evan D.},
        title = "{Isolating the Young Stellar Population in the Outer Disk of NGC 300}",
      journal = {\apj},
         year = 2016,
        month = nov,
       volume = {831},
       number = {2},
          eid = {191},
        pages = {191},
          doi = {10.3847/0004-637X/831/2/191},
archivePrefix = {arXiv},
       eprint = {1609.02106},
 primaryClass = {astro-ph.GA},
       adsurl = {https://ui.adsabs.harvard.edu/abs/2016ApJ...831..191H}
}

@ARTICLE{debattista06,
       author = {{Debattista}, Victor P. and {Mayer}, Lucio and {Carollo}, C. Marcella and {Moore}, Ben and {Wadsley}, James and {Quinn}, Thomas},
        title = "{The Secular Evolution of Disk Structural Parameters}",
      journal = {\apj},
         year = 2006,
        month = jul,
       volume = {645},
       number = {1},
        pages = {209-227},
          doi = {10.1086/504147},
archivePrefix = {arXiv},
       eprint = {astro-ph/0509310},
 primaryClass = {astro-ph},
       adsurl = {https://ui.adsabs.harvard.edu/abs/2006ApJ...645..209D}
}

@ARTICLE{berrier16,
       author = {{Berrier}, Joel C. and {Sellwood}, J.~A.},
        title = "{Mass Distribution and Bar Formation in Growing Disk Galaxy Models}",
      journal = {\apj},
         year = 2016,
        month = nov,
       volume = {831},
       number = {1},
          eid = {65},
        pages = {65},
          doi = {10.3847/0004-637X/831/1/65},
archivePrefix = {arXiv},
       eprint = {1608.04765},
 primaryClass = {astro-ph.GA},
       adsurl = {https://ui.adsabs.harvard.edu/abs/2016ApJ...831...65B}
}

@ARTICLE{mestel63,
       author = {{Mestel}, L.},
        title = "{On the galactic law of rotation}",
      journal = {\mnras},
         year = 1963,
        month = jan,
       volume = {126},
        pages = {553},
          doi = {10.1093/mnras/126.6.553},
       adsurl = {https://ui.adsabs.harvard.edu/abs/1963MNRAS.126..553M}
}

@ARTICLE{dalcanton97,
       author = {{Dalcanton}, Julianne J. and {Spergel}, David N. and {Summers}, F.~J.},
        title = "{The Formation of Disk Galaxies}",
      journal = {\apj},
         year = 1997,
        month = jun,
       volume = {482},
       number = {2},
        pages = {659-676},
          doi = {10.1086/304182},
archivePrefix = {arXiv},
       eprint = {astro-ph/9611226},
 primaryClass = {astro-ph},
       adsurl = {https://ui.adsabs.harvard.edu/abs/1997ApJ...482..659D}
}

@ARTICLE{fall80,
       author = {{Fall}, S.~M. and {Efstathiou}, G.},
        title = "{Formation and rotation of disc galaxies with haloes.}",
      journal = {\mnras},
         year = 1980,
        month = oct,
       volume = {193},
        pages = {189-206},
          doi = {10.1093/mnras/193.2.189},
       adsurl = {https://ui.adsabs.harvard.edu/abs/1980MNRAS.193..189F}
}

@ARTICLE{efstathiou00,
       author = {{Efstathiou}, G.},
        title = "{A model of supernova feedback in galaxy formation}",
      journal = {\mnras},
         year = 2000,
        month = sep,
       volume = {317},
       number = {3},
        pages = {697-719},
          doi = {10.1046/j.1365-8711.2000.03665.x},
archivePrefix = {arXiv},
       eprint = {astro-ph/0002245},
 primaryClass = {astro-ph},
       adsurl = {https://ui.adsabs.harvard.edu/abs/2000MNRAS.317..697E}
}

@ARTICLE{ferguson01,
       author = {{Ferguson}, A.~M.~N. and {Clarke}, C.~J.},
        title = "{The evolution of stellar exponential discs}",
      journal = {\mnras},
         year = 2001,
        month = aug,
       volume = {325},
       number = {2},
        pages = {781-791},
          doi = {10.1046/j.1365-8711.2001.04501.x},
archivePrefix = {arXiv},
       eprint = {astro-ph/0103205},
 primaryClass = {astro-ph},
       adsurl = {https://ui.adsabs.harvard.edu/abs/2001MNRAS.325..781F}
}

@ARTICLE{lin87,
       author = {{Lin}, D.~N.~C. and {Pringle}, J.~E.},
        title = "{The Formation of the Exponential Disk in Spiral Galaxies}",
      journal = {\apjl},
         year = 1987,
        month = sep,
       volume = {320},
        pages = {L87},
          doi = {10.1086/184981},
       adsurl = {https://ui.adsabs.harvard.edu/abs/1987ApJ...320L..87L}
}

@ARTICLE{khop15,
       author = {{Khoperskov}, S.~A. and {Bertin}, G.},
        title = "{Spiral density waves in the outer galactic gaseous discs}",
      journal = {\mnras},
         year = 2015,
        month = aug,
       volume = {451},
       number = {3},
        pages = {2889-2899},
          doi = {10.1093/mnras/stv1145},
archivePrefix = {arXiv},
       eprint = {1505.04598},
 primaryClass = {astro-ph.GA},
       adsurl = {https://ui.adsabs.harvard.edu/abs/2015MNRAS.451.2889K}
}

@ARTICLE{mihos13,
       author = {{Mihos}, J. Christopher and {Harding}, Paul and {Spengler}, Chelsea E. and {Rudick}, Craig S. and {Feldmeier}, John J.},
        title = "{The Extended Optical Disk of M101}",
      journal = {\apj},
         year = 2013,
        month = jan,
       volume = {762},
       number = {2},
          eid = {82},
        pages = {82},
          doi = {10.1088/0004-637X/762/2/82},
archivePrefix = {arXiv},
       eprint = {1211.3095},
 primaryClass = {astro-ph.CO},
       adsurl = {https://ui.adsabs.harvard.edu/abs/2013ApJ...762...82M}
}

@ARTICLE{vandokkum14,
       author = {{van Dokkum}, Pieter G. and {Abraham}, Roberto and {Merritt}, Allison},
        title = "{First Results from the Dragonfly Telephoto Array: The Apparent Lack of a Stellar Halo in the Massive Spiral Galaxy M101}",
      journal = {\apjl},
         year = 2014,
        month = feb,
       volume = {782},
       number = {2},
          eid = {L24},
        pages = {L24},
          doi = {10.1088/2041-8205/782/2/L24},
archivePrefix = {arXiv},
       eprint = {1401.5467},
 primaryClass = {astro-ph.GA},
       adsurl = {https://ui.adsabs.harvard.edu/abs/2014ApJ...782L..24V}
}

@ARTICLE{radburn12,
       author = {{Radburn-Smith}, David J. and {Ro{\v{s}}kar}, Rok and {Debattista}, Victor P. and {Dalcanton}, Julianne J. and {Streich}, David and {de Jong}, Roelof S. and {Vlaji{\'c}}, Marija and {Holwerda}, Benne W. and {Purcell}, Chris W. and {Dolphin}, Andrew E. and {Zucker}, Daniel B.},
        title = "{Outer-disk Populations in NGC 7793: Evidence for Stellar Radial Migration}",
      journal = {\apj},
         year = 2012,
        month = jul,
       volume = {753},
       number = {2},
          eid = {138},
        pages = {138},
          doi = {10.1088/0004-637X/753/2/138},
archivePrefix = {arXiv},
       eprint = {1206.1057},
 primaryClass = {astro-ph.CO},
       adsurl = {https://ui.adsabs.harvard.edu/abs/2012ApJ...753..138R}
}

@ARTICLE{sellwood02,
       author = {{Sellwood}, J.~A. and {Binney}, J.~J.},
        title = "{Radial mixing in galactic discs}",
      journal = {\mnras},
         year = 2002,
        month = nov,
       volume = {336},
       number = {3},
        pages = {785-796},
          doi = {10.1046/j.1365-8711.2002.05806.x},
archivePrefix = {arXiv},
       eprint = {astro-ph/0203510},
 primaryClass = {astro-ph},
       adsurl = {https://ui.adsabs.harvard.edu/abs/2002MNRAS.336..785S}
}

@ARTICLE{roskar08a,
       author = {{Ro{\v{s}}kar}, Rok and {Debattista}, Victor P. and {Stinson}, Gregory S. and {Quinn}, Thomas R. and {Kaufmann}, Tobias and {Wadsley}, James},
        title = "{Beyond Inside-Out Growth: Formation and Evolution of Disk Outskirts}",
      journal = {\apjl},
         year = 2008,
        month = mar,
       volume = {675},
       number = {2},
        pages = {L65},
          doi = {10.1086/586734},
archivePrefix = {arXiv},
       eprint = {0710.5523},
 primaryClass = {astro-ph},
       adsurl = {https://ui.adsabs.harvard.edu/abs/2008ApJ...675L..65R}
}

@ARTICLE{dejong07,
       author = {{de Jong}, Roelof S. and {Seth}, A.~C. and {Radburn-Smith}, D.~J. and {Bell}, E.~F. and {Brown}, T.~M. and {Bullock}, J.~S. and {Courteau}, S. and {Dalcanton}, J.~J. and {Ferguson}, H.~C. and {Goudfrooij}, P. and {Holfeltz}, S. and {Holwerda}, B.~W. and {Purcell}, C. and {Sick}, J. and {Zucker}, D.~B.},
        title = "{Stellar Populations across the NGC 4244 Truncated Galactic Disk}",
      journal = {\apjl},
         year = 2007,
        month = sep,
       volume = {667},
       number = {1},
        pages = {L49-L52},
          doi = {10.1086/522035},
archivePrefix = {arXiv},
       eprint = {0708.0826},
 primaryClass = {astro-ph},
       adsurl = {https://ui.adsabs.harvard.edu/abs/2007ApJ...667L..49D}
}

@ARTICLE{purcell11,
       author = {{Purcell}, Chris W. and {Bullock}, James S. and {Tollerud}, Erik J. and {Rocha}, Miguel and {Chakrabarti}, Sukanya},
        title = "{The Sagittarius impact as an architect of spirality and outer rings in the Milky Way}",
      journal = {\nat},
         year = 2011,
        month = sep,
       volume = {477},
       number = {7364},
        pages = {301-303},
          doi = {10.1038/nature10417},
archivePrefix = {arXiv},
       eprint = {1109.2918},
 primaryClass = {astro-ph.GA},
       adsurl = {https://ui.adsabs.harvard.edu/abs/2011Natur.477..301P}
}

@ARTICLE{struck11,
       author = {{Struck}, Curtis and {Dobbs}, Clare L. and {Hwang}, Jeong-Sun},
        title = "{Slowly breaking waves: the longevity of tidally induced spiral structure}",
      journal = {\mnras},
         year = 2011,
        month = jul,
       volume = {414},
       number = {3},
        pages = {2498-2510},
          doi = {10.1111/j.1365-2966.2011.18568.x},
archivePrefix = {arXiv},
       eprint = {1102.4817},
 primaryClass = {astro-ph.CO},
       adsurl = {https://ui.adsabs.harvard.edu/abs/2011MNRAS.414.2498S}
}

@ARTICLE{putko25,
       author = {{Putko}, Joseph and {Elmegreen}, Bruce G. and {Mu\~noz-Tu\~n\'on}, Casiana and {S\'anchez Almeida}, Jorge and {Akhlaghi}, Mohammad and {Elmegreen}, Debra M. and {Caon}, Nicola},
       title = "{Clumpy, Local Extremely Metal-Poor Galaxies: Photometry and Morphology of the Host Component}",
      journal = {\aap},
         year = 2025,
        month = {},
       volume = {},
       number = {},
        pages = {submitted},
          doi = {},
archivePrefix = {},
       eprint = {},
 primaryClass = {},
       adsurl = {}
}

@ARTICLE{barker12,
       author = {{Barker}, Michael K. and {Ferguson}, Annette M.~N. and {Irwin}, M.~J. and {Arimoto}, N. and {Jablonka}, P.},
        title = "{Quantifying the faint structure of galaxies: the late-type spiral NGC 2403}",
      journal = {\mnras},
         year = 2012,
        month = jan,
       volume = {419},
       number = {2},
        pages = {1489-1506},
          doi = {10.1111/j.1365-2966.2011.19814.x},
archivePrefix = {arXiv},
       eprint = {1109.2625},
 primaryClass = {astro-ph.CO},
       adsurl = {https://ui.adsabs.harvard.edu/abs/2012MNRAS.419.1489B}
}

@ARTICLE{watkins14,
       author = {{Watkins}, Aaron E. and {Mihos}, J. Christopher and {Harding}, Paul and {Feldmeier}, John J.},
        title = "{Searching for Diffuse Light in the M96 Galaxy Group}",
      journal = {\apj},
         year = 2014,
        month = aug,
       volume = {791},
       number = {1},
          eid = {38},
        pages = {38},
          doi = {10.1088/0004-637X/791/1/38},
archivePrefix = {arXiv},
       eprint = {1406.6982},
 primaryClass = {astro-ph.GA},
       adsurl = {https://ui.adsabs.harvard.edu/abs/2014ApJ...791...38W}
}

@ARTICLE{watkins15,
       author = {{Watkins}, Aaron E. and {Mihos}, J. Christopher and {Harding}, Paul},
        title = "{Deep Imaging of M51: a New View of the Whirlpool{\textquoteright}s Extended Tidal Debris}",
      journal = {\apjl},
         year = 2015,
        month = feb,
       volume = {800},
       number = {1},
          eid = {L3},
        pages = {L3},
          doi = {10.1088/2041-8205/800/1/L3},
archivePrefix = {arXiv},
       eprint = {1501.04599},
 primaryClass = {astro-ph.GA},
       adsurl = {https://ui.adsabs.harvard.edu/abs/2015ApJ...800L...3W}
}

@ARTICLE{bernard12,
       author = {{Bernard}, Edouard J. and {Ferguson}, Annette M.~N. and {Barker}, Michael K. and {Irwin}, Michael J. and {Jablonka}, Pascale and {Arimoto}, Nobuo},
        title = "{A deep, wide-field study of Holmberg II with Suprime-Cam: evidence for ram pressure stripping}",
      journal = {\mnras},
         year = 2012,
        month = nov,
       volume = {426},
       number = {4},
        pages = {3490-3500},
          doi = {10.1111/j.1365-2966.2012.22025.x},
archivePrefix = {arXiv},
       eprint = {1208.4808},
 primaryClass = {astro-ph.GA},
       adsurl = {https://ui.adsabs.harvard.edu/abs/2012MNRAS.426.3490B}
}

@ARTICLE{jang20a,
       author = {{Jang}, In Sung and {de Jong}, Roelof S. and {Holwerda}, Benne W. and {Monachesi}, Antonela and {Bell}, Eric F. and {Bailin}, Jeremy},
        title = "{Tracing the anemic stellar halo of M 101}",
      journal = {\aap},
         year = 2020,
        month = may,
       volume = {637},
          eid = {A8},
        pages = {A8},
          doi = {10.1051/0004-6361/201936994},
archivePrefix = {arXiv},
       eprint = {2001.12007},
 primaryClass = {astro-ph.GA},
       adsurl = {https://ui.adsabs.harvard.edu/abs/2020A&A...637A...8J}
}

@ARTICLE{jang20b,
       author = {{Jang}, In Sung and {de Jong}, Roelof S. and {Minchev}, Ivan and {Bell}, Eric F. and {Monachesi}, Antonela and {Holwerda}, Benne W. and {Bailin}, Jeremy and {Smercina}, Adam and {D'Souza}, Richard},
        title = "{Is NGC 300 a pure exponential disk galaxy?}",
      journal = {\aap},
         year = 2020,
        month = aug,
       volume = {640},
          eid = {L19},
        pages = {L19},
          doi = {10.1051/0004-6361/202038651},
archivePrefix = {arXiv},
       eprint = {2007.13749},
 primaryClass = {astro-ph.GA},
       adsurl = {https://ui.adsabs.harvard.edu/abs/2020A&A...640L..19J}
}

@ARTICLE{barton97,
       author = {{Barton}, Ian J. and {Thompson}, Laird A.},
        title = "{Deep Surface Photometry of Spiral Galaxy NGC 5383: Observational Techniques and Halo Constraints}",
      journal = {\aj},
         year = 1997,
        month = aug,
       volume = {114},
        pages = {655-668},
          doi = {10.1086/118500},
       adsurl = {https://ui.adsabs.harvard.edu/abs/1997AJ....114..655B}
}

@ARTICLE{vlajic09,
       author = {{Vlaji{\'c}}, M. and {Bland-Hawthorn}, J. and {Freeman}, K.~C.},
        title = "{The Abundance Gradient in the Extremely Faint Outer Disk of NGC 300}",
      journal = {\apj},
         year = 2009,
        month = may,
       volume = {697},
       number = {1},
        pages = {361-372},
          doi = {10.1088/0004-637X/697/1/361},
archivePrefix = {arXiv},
       eprint = {0903.1855},
 primaryClass = {astro-ph.CO},
       adsurl = {https://ui.adsabs.harvard.edu/abs/2009ApJ...697..361V}
}

@ARTICLE{staudaher19,
       author = {{Staudaher}, Shawn M. and {Dale}, Daniel A. and {van Zee}, Liese},
        title = "{The Extended Disc Galaxy Exploration Science Survey: description and surface brightness profile properties}",
      journal = {\mnras},
         year = 2019,
        month = jun,
       volume = {486},
       number = {2},
        pages = {1995-2010},
          doi = {10.1093/mnras/stz935},
archivePrefix = {arXiv},
       eprint = {1904.00050},
 primaryClass = {astro-ph.GA},
       adsurl = {https://ui.adsabs.harvard.edu/abs/2019MNRAS.486.1995S}
}

@ARTICLE{zhang18,
       author = {{Zhang}, Jielai and {Abraham}, Roberto and {van Dokkum}, Pieter and {Merritt}, Allison and {Janssens}, Steven},
        title = "{The Dragonfly Nearby Galaxies Survey. IV. A Giant Stellar Disk in NGC 2841}",
      journal = {\apj},
         year = 2018,
        month = mar,
       volume = {855},
       number = {2},
          eid = {78},
        pages = {78},
          doi = {10.3847/1538-4357/aaac81},
archivePrefix = {arXiv},
       eprint = {1802.02583},
 primaryClass = {astro-ph.GA},
       adsurl = {https://ui.adsabs.harvard.edu/abs/2018ApJ...855...78Z}
}

@INPROCEEDINGS{abraham17,
       author = {{Abraham}, Roberto and {Merritt}, Allison and {Zhang}, Jielai and {van Dokkum}, Pieter and {Conroy}, Charlie and {Danieli}, Shany and {Mowla}, Lamiya},
        title = "{Probing Galactic Outskirts with Dragonfly}",
    booktitle = {Formation and Evolution of Galaxy Outskirts},
         year = 2017,
       editor = {{Gil de Paz}, Armando and {Knapen}, Johan H. and {Lee}, Janice C.},
       series = {IAU Symposium},
       volume = {321},
        month = mar,
        pages = {137-146},
          doi = {10.1017/S1743921316012291},
       adsurl = {https://ui.adsabs.harvard.edu/abs/2017IAUS..321..137A}
}

@ARTICLE{williams13,
       author = {{Williams}, Benjamin F. and {Dalcanton}, Julianne J. and {Stilp}, Adrienne and {Dolphin}, Andrew and {Skillman}, Evan D. and {Radburn-Smith}, David},
        title = "{The ACS Nearby Galaxy Survey Treasury. XI. The Remarkably Undisturbed NGC 2403 Disk}",
      journal = {\apj},
         year = 2013,
        month = mar,
       volume = {765},
       number = {2},
          eid = {120},
        pages = {120},
          doi = {10.1088/0004-637X/765/2/120},
archivePrefix = {arXiv},
       eprint = {1301.4521},
 primaryClass = {astro-ph.CO},
       adsurl = {https://ui.adsabs.harvard.edu/abs/2013ApJ...765..120W}
}

@ARTICLE{erwin05,
       author = {{Erwin}, Peter and {Beckman}, John E. and {Pohlen}, Michael},
        title = "{Antitruncation of Disks in Early-Type Barred Galaxies}",
      journal = {\apjl},
         year = 2005,
        month = jun,
       volume = {626},
       number = {2},
        pages = {L81-L84},
          doi = {10.1086/431739},
archivePrefix = {arXiv},
       eprint = {astro-ph/0505216},
 primaryClass = {astro-ph},
       adsurl = {https://ui.adsabs.harvard.edu/abs/2005ApJ...626L..81E}
}

@ARTICLE{vlajic11,
       author = {{Vlaji{\'c}}, M. and {Bland-Hawthorn}, J. and {Freeman}, K.~C.},
        title = "{The Structure and Metallicity Gradient in the Extreme Outer Disk of NGC 7793}",
      journal = {\apj},
         year = 2011,
        month = may,
       volume = {732},
       number = {1},
          eid = {7},
        pages = {7},
          doi = {10.1088/0004-637X/732/1/7},
archivePrefix = {arXiv},
       eprint = {1101.0607},
 primaryClass = {astro-ph.GA},
       adsurl = {https://ui.adsabs.harvard.edu/abs/2011ApJ...732....7V}
}

@ARTICLE{quillen09,
       author = {{Quillen}, A.~C. and {Minchev}, Ivan and {Bland-Hawthorn}, Joss and {Haywood}, Misha},
        title = "{Radial mixing in the outer Milky Way disc caused by an orbiting satellite}",
      journal = {\mnras},
         year = 2009,
        month = aug,
       volume = {397},
       number = {3},
        pages = {1599-1606},
          doi = {10.1111/j.1365-2966.2009.15054.x},
archivePrefix = {arXiv},
       eprint = {0903.1851},
 primaryClass = {astro-ph.GA},
       adsurl = {https://ui.adsabs.harvard.edu/abs/2009MNRAS.397.1599Q}
}

@ARTICLE{atkinson13,
       author = {{Atkinson}, Adam M. and {Abraham}, Roberto G. and {Ferguson}, Annette M.~N.},
        title = "{Faint Tidal Features in Galaxies within the Canada-France-Hawaii Telescope Legacy Survey Wide Fields}",
      journal = {\apj},
         year = 2013,
        month = mar,
       volume = {765},
       number = {1},
          eid = {28},
        pages = {28},
          doi = {10.1088/0004-637X/765/1/28},
archivePrefix = {arXiv},
       eprint = {1301.4275},
 primaryClass = {astro-ph.CO},
       adsurl = {https://ui.adsabs.harvard.edu/abs/2013ApJ...765...28A}
}

@ARTICLE{martin12,
       author = {{Mart{\'\i}n-Navarro}, Ignacio and {Bakos}, Judit and {Trujillo}, Ignacio and {Knapen}, Johan H. and {Athanassoula}, E. and {Bosma}, Albert and {Comer{\'o}n}, S{\'e}bastien and {Elmegreen}, Bruce G. and {Erroz-Ferrer}, Santiago and {Gadotti}, Dimitri A. and {Gil de Paz}, Armando and {Hinz}, Joannah L. and {Ho}, Luis C. and {Holwerda}, Benne W. and {Kim}, Taehyun and {Laine}, Jarkko and {Laurikainen}, Eija and {Men{\'e}ndez-Delmestre}, Kar{\'\i}n. and {Mizusawa}, Trisha and {Mu{\~n}oz-Mateos}, Juan-Carlos and {Regan}, Michael W. and {Salo}, Heikki and {Seibert}, Mark and {Sheth}, Kartik},
        title = "{A unified picture of breaks and truncations in spiral galaxies from SDSS and S$^{4}$G imaging}",
      journal = {\mnras},
         year = 2012,
        month = dec,
       volume = {427},
       number = {2},
        pages = {1102-1134},
          doi = {10.1111/j.1365-2966.2012.21929.x},
archivePrefix = {arXiv},
       eprint = {1208.2893},
 primaryClass = {astro-ph.CO},
       adsurl = {https://ui.adsabs.harvard.edu/abs/2012MNRAS.427.1102M}
}

@ARTICLE{bournaud07,
       author = {{Bournaud}, Fr{\'e}d{\'e}ric and {Elmegreen}, Bruce G. and {Elmegreen}, Debra Meloy},
        title = "{Rapid Formation of Exponential Disks and Bulges at High Redshift from the Dynamical Evolution of Clump-Cluster and Chain Galaxies}",
      journal = {\apj},
         year = 2007,
        month = nov,
       volume = {670},
       number = {1},
        pages = {237-248},
          doi = {10.1086/522077},
archivePrefix = {arXiv},
       eprint = {0708.0306},
 primaryClass = {astro-ph},
       adsurl = {https://ui.adsabs.harvard.edu/abs/2007ApJ...670..237B}
}

@ARTICLE{struck17,
       author = {{Struck}, Curtis and {Elmegreen}, Bruce G.},
        title = "{Exponential profiles from stellar scattering off of interstellar clumps and holes in dwarf galaxy discs}",
      journal = {\mnras},
         year = 2017,
        month = jul,
       volume = {469},
       number = {1},
        pages = {1157-1165},
          doi = {10.1093/mnras/stx918},
archivePrefix = {arXiv},
       eprint = {1704.03831},
 primaryClass = {astro-ph.GA},
       adsurl = {https://ui.adsabs.harvard.edu/abs/2017MNRAS.469.1157S}
}

@ARTICLE{putko19,
       author = {{Putko}, J. and {S{\'a}nchez Almeida}, J. and {Mu{\~n}oz-Tu{\~n}{\'o}n}, C. and {Asensio Ramos}, A. and {Elmegreen}, B.~G. and {Elmegreen}, D.~M.},
        title = "{Inferring the 3D Shapes of Extremely Metal-poor Galaxies from Sets of Projected Shapes}",
      journal = {\apj},
         year = 2019,
        month = sep,
       volume = {883},
       number = {1},
          eid = {10},
        pages = {10},
          doi = {10.3847/1538-4357/ab365a},
archivePrefix = {arXiv},
       eprint = {1907.10496},
 primaryClass = {astro-ph.GA},
       adsurl = {https://ui.adsabs.harvard.edu/abs/2019ApJ...883...10P}
}

@ARTICLE{elmegreen13,
       author = {{Elmegreen}, Bruce G. and {Struck}, Curtis},
        title = "{Exponential Galaxy Disks from Stellar Scattering}",
      journal = {\apjl},
         year = 2013,
        month = oct,
       volume = {775},
       number = {2},
          eid = {L35},
        pages = {L35},
          doi = {10.1088/2041-8205/775/2/L35},
archivePrefix = {arXiv},
       eprint = {1308.5236},
 primaryClass = {astro-ph.GA},
       adsurl = {https://ui.adsabs.harvard.edu/abs/2013ApJ...775L..35E}
}

@ARTICLE{elmegreen16,
       author = {{Elmegreen}, Bruce G. and {Struck}, Curtis},
        title = "{Exponential Disks from Stellar Scattering. III. Stochastic Models}",
      journal = {\apj},
         year = 2016,
        month = oct,
       volume = {830},
       number = {2},
          eid = {115},
        pages = {115},
          doi = {10.3847/0004-637X/830/2/115},
archivePrefix = {arXiv},
       eprint = {1607.07595},
 primaryClass = {astro-ph.GA},
       adsurl = {https://ui.adsabs.harvard.edu/abs/2016ApJ...830..115E}
}

@ARTICLE{wu20,
       author = {{Wu}, Jian and {Struck}, Curtis and {D'Onghia}, Elena and {Elmegreen}, Bruce G.},
        title = "{Stellar scattering and the formation of exponential discs in self-gravitating systems}",
      journal = {\mnras},
         year = 2020,
        month = dec,
       volume = {499},
       number = {2},
        pages = {2672-2684},
          doi = {10.1093/mnras/staa2750},
archivePrefix = {arXiv},
       eprint = {2009.01929},
 primaryClass = {astro-ph.GA},
       adsurl = {https://ui.adsabs.harvard.edu/abs/2020MNRAS.499.2672W}
}

@ARTICLE{wu22,
       author = {{Wu}, Jian and {Struck}, Curtis and {Elmegreen}, Bruce G. and {D'Onghia}, Elena},
        title = "{Orbits and action changes during star-clump encounters responsible for the origin of exponential discs in dwarf galaxies}",
      journal = {\mnras},
         year = 2022,
        month = dec,
       volume = {517},
       number = {3},
        pages = {4417-4435},
          doi = {10.1093/mnras/stac2870},
archivePrefix = {arXiv},
       eprint = {2210.00651},
 primaryClass = {astro-ph.GA},
       adsurl = {https://ui.adsabs.harvard.edu/abs/2022MNRAS.517.4417W}
}

@ARTICLE{wu23,
       author = {{Wu}, Jian and {Struck}, Curtis and {Elmegreen}, Bruce G. and {D'Onghia}, Elena},
        title = "{Exponential galaxy discs as the quasi-stationary distribution in a Markov chain model simulating stellar scattering}",
      journal = {\mnras},
         year = 2023,
        month = jul,
       volume = {522},
       number = {3},
        pages = {3948-3964},
          doi = {10.1093/mnras/stad1280},
archivePrefix = {arXiv},
       eprint = {2304.11774},
 primaryClass = {astro-ph.GA},
       adsurl = {https://ui.adsabs.harvard.edu/abs/2023MNRAS.522.3948W}
}

@ARTICLE{schonrich09,
       author = {{Sch{\"o}nrich}, Ralph and {Binney}, James},
        title = "{Chemical evolution with radial mixing}",
      journal = {\mnras},
         year = 2009,
        month = jun,
       volume = {396},
       number = {1},
        pages = {203-222},
          doi = {10.1111/j.1365-2966.2009.14750.x},
archivePrefix = {arXiv},
       eprint = {0809.3006},
 primaryClass = {astro-ph},
       adsurl = {https://ui.adsabs.harvard.edu/abs/2009MNRAS.396..203S}
}

@ARTICLE{vijay25,
       author = {{Vijayakumar}, Vivek and {Sun}, Jiayi and {Ostriker}, Eve C. and {Di Teodoro}, Enrico M. and {Haubner}, Konstantin and {Kim}, Chang-Goo and {Leroy}, Adam K. and {Querejeta}, Miguel},
        title = "{Modeling the Mass Distribution and Gravitational Potential of Nearby Disk Galaxies: Implications for the ISM Dynamical Equilibrium}",
      journal = {arXiv e-prints},
         year = 2025,
        month = jun,
          eid = {arXiv:2506.22381},
        pages = {arXiv:2506.22381},
          doi = {10.48550/arXiv.2506.22381},
archivePrefix = {arXiv},
       eprint = {2506.22381},
 primaryClass = {astro-ph.GA},
       adsurl = {https://ui.adsabs.harvard.edu/abs/2025arXiv250622381V}
}

@ARTICLE{wang19,
       author = {{Wang}, Enci and {Lilly}, Simon J. and {Pezzulli}, Gabriele and {Matthee}, Jorryt},
        title = "{On the Elevation and Suppression of Star Formation within Galaxies}",
      journal = {\apj},
         year = 2019,
        month = jun,
       volume = {877},
       number = {2},
          eid = {132},
        pages = {132},
          doi = {10.3847/1538-4357/ab1c5b},
archivePrefix = {arXiv},
       eprint = {1901.10276},
 primaryClass = {astro-ph.GA},
       adsurl = {https://ui.adsabs.harvard.edu/abs/2019ApJ...877..132W}
}

@ARTICLE{casasola17,
       author = {{Casasola}, V. and {Cassar{\`a}}, L.~P. and {Bianchi}, S. and {Verstocken}, S. and {Xilouris}, E. and {Magrini}, L. and {Smith}, M.~W.~L. and {De Looze}, I. and {Galametz}, M. and {Madden}, S.~C. and {Baes}, M. and {Clark}, C. and {Davies}, J. and {De Vis}, P. and {Evans}, R. and {Fritz}, J. and {Galliano}, F. and {Jones}, A.~P. and {Mosenkov}, A.~V. and {Viaene}, S. and {Ysard}, N.},
        title = "{Radial distribution of dust, stars, gas, and star-formation rate in DustPedia face-on galaxies}",
      journal = {\aap},
         year = 2017,
        month = sep,
       volume = {605},
          eid = {A18},
        pages = {A18},
          doi = {10.1051/0004-6361/201731020},
archivePrefix = {arXiv},
       eprint = {1706.05351},
 primaryClass = {astro-ph.GA},
       adsurl = {https://ui.adsabs.harvard.edu/abs/2017A&A...605A..18C}
}

@ARTICLE{lin24,
       author = {{Lin}, Lin and {Shen}, Shiyin and {Yesuf}, Hassen M. and {Mao}, Ye-Wei and {Hao}, Lei},
        title = "{Radial Profiles of {\ensuremath{\Sigma}}$_{*}$, {\ensuremath{\Sigma}}$_{SFR}$, Gas Metallicity, and Their Correlations across the Galactic Mass{\textendash}Size Plane}",
      journal = {\apj},
         year = 2024,
        month = dec,
       volume = {977},
       number = {2},
          eid = {175},
        pages = {175},
          doi = {10.3847/1538-4357/ad8a61},
archivePrefix = {arXiv},
       eprint = {2410.16651},
 primaryClass = {astro-ph.GA},
       adsurl = {https://ui.adsabs.harvard.edu/abs/2024ApJ...977..175L}
}

@ARTICLE{gonzalez16,
       author = {{Gonz{\'a}lez Delgado}, R.~M. and {Cid Fernandes}, R. and {P{\'e}rez}, E. and {Garc{\'\i}a-Benito}, R. and {L{\'o}pez Fern{\'a}ndez}, R. and {Lacerda}, E.~A.~D. and {Cortijo-Ferrero}, C. and {de Amorim}, A.~L. and {Vale Asari}, N. and {S{\'a}nchez}, S.~F. and {Walcher}, C.~J. and {Wisotzki}, L. and {Mast}, D. and {Alves}, J. and {Ascasibar}, Y. and {Bland-Hawthorn}, J. and {Galbany}, L. and {Kennicutt}, R.~C. and {M{\'a}rquez}, I. and {Masegosa}, J. and {Moll{\'a}}, M. and {S{\'a}nchez-Bl{\'a}zquez}, P. and {V{\'\i}lchez}, J.~M.},
        title = "{Star formation along the Hubble sequence. Radial structure of the star formation of CALIFA galaxies}",
      journal = {\aap},
         year = 2016,
        month = may,
       volume = {590},
          eid = {A44},
        pages = {A44},
          doi = {10.1051/0004-6361/201628174},
archivePrefix = {arXiv},
       eprint = {1603.00874},
 primaryClass = {astro-ph.GA},
       adsurl = {https://ui.adsabs.harvard.edu/abs/2016A&A...590A..44G}
}

@ARTICLE{xu24,
       author = {{Xu}, Dewang and {Yu}, Si-Yue},
        title = "{JWST reveals a high fraction of disk breaks at 1 {\ensuremath{\leq}} z {\ensuremath{\leq}} 3}",
      journal = {\aap},
         year = 2024,
        month = feb,
       volume = {682},
          eid = {L17},
        pages = {L17},
          doi = {10.1051/0004-6361/202449252},
archivePrefix = {arXiv},
       eprint = {2402.04233},
 primaryClass = {astro-ph.GA},
       adsurl = {https://ui.adsabs.harvard.edu/abs/2024A&A...682L..17X}
}

@ARTICLE{zheng15,
       author = {{Zheng}, Zheng and {Thilker}, David A. and {Heckman}, Timothy M. and {Meurer}, Gerhardt R. and {Burgett}, W.~S. and {Chambers}, K.~C. and {Huber}, M.~E. and {Kaiser}, N. and {Magnier}, E.~A. and {Metcalfe}, N. and {Price}, P.~A. and {Tonry}, J.~L. and {Wainscoat}, R.~J. and {Waters}, C.},
        title = "{The Structure and Stellar Content of the Outer Disks of Galaxies: A New View from the Pan-STARRS1 Medium Deep Survey}",
      journal = {\apj},
         year = 2015,
        month = feb,
       volume = {800},
       number = {2},
          eid = {120},
        pages = {120},
          doi = {10.1088/0004-637X/800/2/120},
archivePrefix = {arXiv},
       eprint = {1412.3209},
 primaryClass = {astro-ph.GA},
       adsurl = {https://ui.adsabs.harvard.edu/abs/2015ApJ...800..120Z}
}

@ARTICLE{bakos08,
       author = {{Bakos}, Judit and {Trujillo}, Ignacio and {Pohlen}, Michael},
        title = "{Color Profiles of Spiral Galaxies: Clues on Outer-Disk Formation Scenarios}",
      journal = {\apjl},
         year = 2008,
        month = aug,
       volume = {683},
       number = {2},
        pages = {L103},
          doi = {10.1086/591671},
archivePrefix = {arXiv},
       eprint = {0807.2776},
 primaryClass = {astro-ph},
       adsurl = {https://ui.adsabs.harvard.edu/abs/2008ApJ...683L.103B}
}

@ARTICLE{spitzer51,
       author = {{Spitzer}, Jr., Lyman and {Schwarzschild}, Martin},
        title = "{The Possible Influence of Interstellar Clouds on Stellar Velocities.}",
      journal = {\apj},
         year = 1951,
        month = nov,
       volume = {114},
        pages = {385},
          doi = {10.1086/145478},
       adsurl = {https://ui.adsabs.harvard.edu/abs/1951ApJ...114..385S}
}
\bibliographystyle{aasjournal}

%

\end{document}